\documentclass{aa}  

\usepackage{booktabs}

\usepackage{float}

\usepackage{graphicx}

\usepackage{txfonts}

\usepackage{natbib}
\bibpunct{(}{)}{;}{a}{}{,}

\usepackage{booktabs}
\usepackage{makecell}
\usepackage{color, colortbl}

\usepackage[colorlinks=true, linkcolor=blue, citecolor=blue]{hyperref}

\usepackage[normalem]{ulem}
\usepackage{xcolor}

\newcommand{\av}[1]{\left\langle #1\right\rangle}

\usepackage{csquotes}

\usepackage{soul}

\begin{document}

   \title{Non-thermal plasma density redistribution in planetary magnetospheres due to ion-cyclotron waves}

   \author{Joaquín Espinoza-Troni\inst{1}
          \and
          Felipe A Asenjo \inst{2} 
          \and 
          Pablo S Moya\inst{1}
          }

\institute{Departamento de Física, Facultad de Ciencias,
Universidad de Chile, Santiago, Chile\\
\email{joaquinun@gmail.com, pablo.moya@uchile.cl}
\and
Facultad de Ingenier\'ia y Ciencias,
Universidad Adolfo Ib\'a\~nez, Santiago, Chile. \\
\email{felipe.asenjo@uai.cl}
}

\date{}

\abstract
{Planetary magnetospheres throughout the solar system exhibit diverse plasma environments where Ultra-low frequency (ULF) pulsations can induce nonlinear ponderomotive effects. Recent analytical studies have predicted a significant influence of the Kappa velocity distribution on the ponderomotive force (PF) induced by electromagnetic ion cyclotron (EMIC) waves. Since suprathermal populations modeled by Kappa distributions are ubiquitous in planetary magnetospheres (from Mercury to the Ice Giants), their influence on ponderomotive phenomena must be accounted for.}
{We investigate the field-aligned plasma density redistribution driven by the PF of traveling EMIC waves, performing a comparative analysis across the different regimes characteristic of the planetary magnetospheres of our solar system.}
{We apply a generalized slow-time-scale force balance equation to model the stationary plasma density solutions for each planetary magnetosphere. The model incorporates the PF induced by traveling EMIC waves in low-beta plasmas ($\beta \ll 1$) with isotropic Kappa distributions. To enable a systematic comparison, the wave modulation is described using the WKB approximation in a dipole magnetic field model, neglecting curvature effects to first order in planetary fields.}
{We find that the plasma response varies significantly depending on the magnetospheric parameters: a decrease in the kappa parameter and an increase in plasma beta counteract plasma accumulation towards the equator. In low-beta planetary environments, non-thermal effects significantly reduce the nonlinear response to short-period pulsations without altering the qualitative behavior predicted by Maxwellian models. We also characterize how the critical parameter governing the phase transition between equatorial density minima and maxima varies with the specific combinations of plasma beta, the kappa parameter, and L-shell observed across the solar system. Furthermore, applying this model to the dayside, low-beta inner magnetospheres of Earth, Jupiter, and Saturn reveals that the PF exerts a substantial influence primarily during exceptionally high-amplitude EMIC wave events.}
{Our analytical study demonstrates that the non-thermal properties of plasma are a governing factor in the field-aligned density redistribution driven by ULF waves. These results highlight the necessity of incorporating non-thermal effects to accurately model ponderomotive phenomena in the multifaceted types of planetary magnetospheres of the solar system.}


   \maketitle

\section{Introduction}

Planetary magnetospheres throughout the solar system act as cavities that support a rich variety of Ultra-low frequency (ULF) waves. While most studies focus on the Earth's magnetosphere, ULF waves have been observed extensively in the magnetospheres of other planets. Mercury's magnetosphere exhibits ULF waves in the ion cyclotron frequency range, observed by the Mariner 10 \citep{Russell_1989,Kim_etal_2015} and MESSENGER \citep{Boardsen_etal_2009, Boardsen_etal_2009b,Boardsen_etal_2012} spacecraft missions. Farther out, EMIC-mode ULF waves were detected at Jupiter by the Ulysses spacecraft \citep{Petkaki_Dougherty_2001}, and, more recently, a global statistical analysis of Juno data has further characterized Jovian ULF waves \citep{Sun_Xie_etal_2024}. At Saturn, Voyager 1, 2 \citep{Orlowski_etal_1992}, and Cassini \citep{Kleindienst_etal_2009,Andres_etal_2013} spacecraft detected ULF waves, while Voyager 2 observations also revealed ULF activity at the Ice Giants, Uranus and Neptune \citep{Russell_and_Lepping_1992,Farrell_1993}.

Although the underlying physics is shared, planetary magnetospheres differ in size, structure, and composition, affecting ULF propagation. For instance, Jupiter and Saturn exhibit longer-period pulsations than Earth due to their larger magnetospheres. Indeed, the ULF range can be extended here, passing from 0.2--600~s at Earth to tens of minutes for both Jupyter and Saturn (supporting frequencies of $< 1\,\text{mHz}$) \citep{Delamere_2016}. Historically, these waves have been classified according to terrestrial standards. They correspond to geomagnetic pulsations spanning frequencies from the lowest that the Earth's magnetospheric cavity can support up to various ion gyro-frequencies, corresponding to a range from 0.001--5~Hz \citep{allan_1992,Hughes_Jeffrey_1994, Hartinger_2022}. These are classified into five groups based on their period: Pc1 (0.2--5~s), Pc2 (5--10~s), Pc3 (10--45~s), Pc4 (45--150~s), and Pc5 (150--600~s) \citep{Hughes_Jeffrey_1994}.

These perturbations can also be broadly classified into long-period and short-period pulsations \citep{Kangas_Guglielmi_Pokhotelov_1998}. Long-period pulsations span the lower end of the frequency band, possessing wavelengths comparable to the size of Earth's magnetosphere, and are often well described using a magnetohydrodynamic (MHD) approximation. In contrast, short-period pulsations (Pc1 and Pc2), typically observed as electromagnetic ion cyclotron (EMIC) traveling waves, are better described by other mathematical approximations, such as multi-fluid or kinetic formalisms \citep{Mursula_Braysy_Niskala_Russell_2001}.

Beyond their classification, ULF waves take part in several magnetospheric phenomena. Indeed, extensive research has been conducted in the last decades on the influence of ULF pulsations on the background magnetospheric plasma due to the ponderomotive force (PF) \citep{Lundin_Guglielmi_2007,Lundin_and_Lidgren_2022}, a time-averaged nonlinear force arising from the interaction of quasi-monochromatic or spatially inhomogeneous waves with plasma \citep{Kentwell_1987}.

The PF plays an essential role in the physics of ULF waves and significantly contributes to plasma density redistribution in planetary magnetospheres. It provides a useful framework for studying complex wave-plasma interactions on slow timescales and has frequently been invoked to describe a wide range of phenomena. For instance, this nonlinear effect can partially contribute to plasma acceleration in the polar cusps through the action of traveling Pc1-2 Alfvén and EMIC waves \citep{Guglielmi_and_Lundin_2001}, generate ambipolar electric fields impacting heavy ion acceleration \citep{Li_and_Temerin_1993, Guglielmi_and_Feygin_2023, Guglielmi_and_Feygin_2024}, and potentially influence solar wind deceleration in the foreshock \citep{Guglielmi_Potapov_Feygin_2025}. Most theoretical work has focused on Earth, investigating effects from cavity modes to toroidal eigenmodes in simplified dipole models \citep{Allan_etal_1991, Allan_1993, Allan_1994, Guglielmi_Hayashi_Lundin_Potapov_1999} or non-dipolar configurations \citep{Antonova_Shabanskii_1968, Nekrasov_Feygin_2012, Nekrasov_Feygin_2015, Nekrasov_Feygin_2016, Nekrasov_Feygin_2018}, and correlating with observational features like density clouds \citep{Chappell_1974, Anderson_etal_1992} or IMF orientation \citep{Guglielmi_Feygin_2018}. However, the universality of these waves suggests similar processes occur elsewhere in the solar system.

Understanding the impact of ULF waves in these diverse environments requires considering the specific plasma properties of each planet. Due to low collision rates, most space plasmas deviate from thermal equilibrium and are well described by the family of Kappa distributions \citep{Lazar_etal_2023}. These distributions depend on the spectral index $\kappa$, allowing the modeling of suprathermal tails widely observed in near-Earth space environments \citep{lazar_kappa_2021}. Such suprathermal tails are ubiquitous in the solar system; they have been observed and modeled with power law or Kappa distributions in the magnetosphere of Jupiter \citep{Mauk_etal_1996, Mauk_etal_2004}, Saturn \citep{Krimigis_etal_1983,Dialynas_etal_2009}, Mercury \citep{Christon_1987}, Uranus \citep{Krimigis_etal_1986, Mauk_1987}, and Neptune \citep{Krimigis_etal_1989, Mauk_etal_1991}, with kappa values that can vary between 3--10 depending on the magnetospheric zone and the planet \citep{Kirpichev_2020}. Recent analytical studies have highlighted the influence of these non-thermal effects on the PF. \citet{Espinoza-Troni_2023} showed that, in low-temperature unmagnetized plasmas with isotropic Kappa distributions, the $\kappa$ parameter significantly affects the temporal term of the PF due to transverse waves. \citet{Espinoza-Troni_etal_2024} extended these results to low-temperature magnetized plasmas, finding that for all terms of the PF, the impact of the Kappa distribution cannot always be neglected, and increases with the plasma beta.

In this work, we extend previous studies of plasma density redistribution driven by the PF of traveling EMIC waves in non-thermal plasmas, expanding the context to analyze and compare different planetary magnetospheres. We employ the PF expressions from \citet{Espinoza-Troni_etal_2024}, which assume field-aligned propagation of EMIC waves in low-temperature isotropic Kappa plasmas. A dipole approximation of planetary magnetospheres is adopted, enabling a first comparative assessment across the solar system. Although rough for non-dipolar configurations such as Neptune and Uranus \citep{Holme_etal_1996}, this approach provides a useful first-order assessment of non-thermal effects on plasma redistribution driven by the PF in the solar system.

This article is structured as follows: Section \ref{sec:2} reviews the low-time-scale fluid equations of the plasma and introduces the notation and normalization used throughout. Section \ref{sec:3} provides a brief overview of the PF and its different terms. Section \ref{sec:4} presents the ion dispersion relation of EMIC waves for isotropic Kappa plasmas in low-beta conditions. Section \ref{sec:5} computes the different PF terms. To this end, the magnitude of the wave's electric field is determined as a function of the background variables by solving the wave equation in the WKB approximation. Section \ref{sec:6} presents numerical solutions for the differential equation of the plasma density redistribution, and analyzes its behavior for different values of the plasma beta, the L-shell, and the wave's frequency in the context of the planetary magnetospheres of our solar system. Section \ref{sec:7} analyzes how the magnitude of the ponderomotive effect depends on the relative wave amplitude and the average magnetospheric conditions typically observed across the planetary magnetospheres in our solar system. Finally, Section \ref{sec:8} summarizes the main conclusions of this work.

\section{Slow-time-scale one-fluid equation}\label{sec:2}

In this section, we review the equations governing the slow-time-scale dynamics of plasma under the influence of fast-time-scale electromagnetic waves in the magnetospheres of generic planets. The slow-time-scale component of plasma quantities is defined as the low-pass filtered version of the quantity with respect to the wave's carrier frequency. Accordingly, we introduce the time-average operator as
\begin{equation}
    \av{A} = \frac{1}{\tau}\int_t^{t+\tau} A dt,
\end{equation}
where $A$ is a plasma quantity and $\tau \gg 2\pi/\omega$, with $\omega$ the carrier frequency of the fluctuations. Plasma variables can then be decomposed into slow and fast-time-scale components,
\begin{equation}
    A = \av{A} + \tilde{A},
\label{eq:Time-scale_separation}
\end{equation}
where $\tilde{A}$ denotes the fast-time scale contribution. We also define the following scale parameters 
\begin{equation}
    \frac{|\tilde{A}|}{A_0} \sim \zeta, \qquad   \frac{1}{|\av{A}|}\left|\frac{\partial \av{A}}{\partial t}\right| \sim \lambda \omega, \qquad     \frac{1}{|\av{A}|}\left|\nabla \av{A}\right| \sim \eta k
\label{eq:scale_parameters}
\end{equation}
where $\zeta \ll 1$, $\lambda \ll 1$, and $\eta \ll 1$ are small parameters, $A_0$ is the characteristic large-scale value of the corresponding variable, and $k$ is the wavenumber of the fluctuations. Within this framework, the fast-time-scale terms are treated as first-order perturbations, while the slow-time-scale quantities are assumed to vary only weakly in space and time over a wave period. By construction, the fast-time-scale fluctuations average to zero, $\av{\tilde{A}} = 0.$

This time-scale separation (Eq. \ref{eq:Time-scale_separation}) can be applied to the plasma fluid and Maxwell equations and then averaged to obtain the slow-time-scale fluid equations of the plasma \citep{Tskhakaya_1981,Karpman_Shagalov_1982,Lee_Nam_Parks_1983}. The fast-time-scale equations are recovered by subtracting the slow-time-scale equations from the full system, under the approximation $\av{\tilde{f}\tilde{g}} -\tilde{f}\tilde{g} \backsimeq 0$, and retaining terms up to second order in $\tilde{A}$. With these considerations, the slow-time-scale momentum equation becomes 
\begin{equation}
\begin{split}
    & m_\alpha \av{n_\alpha} \left[ \frac{\partial \mathbf{u}_\alpha}{\partial t} + \mathbf{u}_\alpha \cdot \nabla \mathbf{u}_\alpha \right] = q_\alpha\av{n_\alpha}\av{\mathbf{E}} \\
    & + \frac{q_\alpha \av{n_\alpha}}{c}\mathbf{u}_\alpha\times\av{\mathbf{B}} -\nabla\cdot \av{\mathbf{P}_\alpha} + m_\alpha \av{n_\alpha}\mathbf{g} + \mathbf{f}_\alpha, 
\end{split}
\end{equation}
where the subscript $\alpha$ denotes the species, $n_\alpha$ is the number density, $\mathbf{B}$ is the magnetic field, $\mathbf{E}$ is the electric field, $\mathbf{g}$ is the gravity acceleration of the planet (non-constant), and $\mathbf{u}_\alpha$ is the renormalized slow-time-scale fluid velocity, redefined for consistency with the continuity equation. Different renormalizations provide different interpretations of $\mathbf{u}_\alpha$ as discussed in \citet{Kentwell_1987}. Still, since we consider stationary electromagnetic field wave amplitudes and the PF along the geomagnetic field (see below), this choice does not affect our results (see Eqs. (30) and (34) of \citet{Karpman_Shagalov_1982}). The PF  $\mathbf{f}_\alpha$ encompasses the correlation effects of the fast-time scale quantities due to the electromagnetic field perturbation in the slow-time scale dynamics. In the next section, we review its expression within the Washimi-Karpman formalism. 

We consider a quasi-neutral plasma composed of electrons and ions with isotropic pressure tensors. Assuming force balance along the background magnetic field lines, we obtain
\begin{equation}
    0 = -\nabla_\parallel p + M n g_\parallel + f_\parallel,
\end{equation}
where $\nabla_\parallel = \mathbf{\hat{b}}\cdot \nabla$, $g_\parallel = \mathbf{\hat{b}}\cdot \mathbf{g}$, $f_\parallel = \mathbf{\hat{b}}\cdot (\mathbf{f}_e+\mathbf{f}_i)$, $M = m_i + m_e \approx m_i$, $n = \av{n_e} = \av{n_i}$ and $\mathbf{\hat{b}} = \av{\mathbf{B}}/|\av{\mathbf{B}}|$. We take the electrons' and ions' temperatures to be equal $T=T_i = T_e$, with $T$ the temperature in the Maxwellian limit. For the pressure closure, we adopt \citep{Lazar_etal_2017}
\begin{equation}
p = \left(\frac{\kappa}{\kappa-3/2}\right)2k_B T n = \left(\frac{\kappa}{\kappa-3/2}\right)(\beta_{0}c_{A0}^2Mn_0) \bar{n},
\end{equation}
where $k_B$ is the Boltzmann constant, $n_0 = n(x_\parallel = 0)$ is the equatorial number density at the equator (i.e at $x_\parallel = 0$), with $x_\parallel$ the distance along the geomagnetic field lines, $\bar{n}=n/n_0$ is the normalized number density, $\beta_{\kappa 0} = [\kappa-(\kappa-3/2)]\beta_{0}$ with $\beta_{0} = 8\pi n_0 k_B T/B_0^2$ the plasma beta at the equator, and $B_0 = |\av{\mathbf{B}(x_\parallel = 0)}|$  the magnitude of the background equatorial magnetic field. The kappa parameter $\kappa$ quantifies deviations from the thermal equilibrium. If we normalize the directional derivative with the planet radius $R_p$, i.e, $\bar{\nabla}_\parallel = R_p \nabla_\parallel$, the force balance equation becomes
\begin{equation}
    0 = -\beta_{\kappa 0}\bar{\nabla}_\parallel \bar{n} + C_g \bar{n} \bar{g}_\parallel + \nu^2 \bar{f}_\parallel,
\label{eq:Force_balance_normalized}
\end{equation}
where the PF is normalized as  $\bar{f_\parallel} =  {4\pi R_p}  f_\parallel/({B_0^2 \nu^2})$, with $\nu$ as the ratio between the wave amplitude and the background equatorial magnetic field (to be defined more precisely later). The gravity acceleration is normalized as $\bar{g}_\parallel = g_\parallel/g_0$ with $g_0 = GM_p/R_p^2$ the gravity of the planet on its surface, with $G$ the universal gravity constant, and $M_p$ the mass of the planet. We also define the dimensionless parameter $C_g = (g_0R_p/c^2)(c/c_{A0})^2$, which accounts for gravitational effects, with $c$ the speed of light and $c_{A0} = B_0/\sqrt{4\pi M n_0}$ the equatorial Alfvén velocity. 

To calculate the PF, we also require an expression for the field-aligned spatial variation of the fast-time-scale electromagnetic field. From the Maxwell equations, the wave equation for the fast-time-scale electric field is 
\begin{equation}
    4\pi \frac{\partial \tilde{\mathbf{J}}}{\partial t} = c^2\nabla^2 \tilde{\mathbf{E}} - c^2\nabla (\nabla \cdot \tilde{\mathbf{E}})-\frac{\partial^2\tilde{\mathbf{E}}}{\partial t^2},
\label{eq:WaveEq}
\end{equation}
where $\mathbf{\tilde{J}}$ and $\mathbf{\tilde{E}}$ are the fast-time-scale current density and electric field, respectively \citep{Lee_Nam_Parks_1998}.

\section{Washimi-Karpman ponderomotive force}\label{sec:3}

The PF is a nonlinear phenomenon arising from the interaction of a high-frequency field (relative to the slow background) with the plasma, affecting its slow-time-scale dynamics \citep{Kentwell_1987}. In what follows, we adopt the Washimi-Karpman PF formalism \citep{Washimi_Karpman_1976,Karpman_Shagalov_1982}.  

In this framework, the fast-time scale electromagnetic field is represented as a quasi-monochromatic wave $\mathbf{\tilde{E}} = (1/2)\left[\mathbf{\hat{E}}(\mathbf{r},t)e^{-i\omega t} + \mathbf{\hat{E}}^*(\mathbf{r},t)e^{i\omega t} \right]$, where $|\hat{\mathbf{E}}(\mathbf{r},t)|$ varies slowly in space and time compared to the wave carrier frequency and wavenumber. In the presence of a background magnetic field, the Washimi-Karpman PF $\mathbf{f} = \sum_\alpha \mathbf{f}_\alpha$ consists of four contributions. A term $\mathbf{f}_{(s)}$ associated with the spatial modulation of the electric field magnitude, a term $\mathbf{f}_{(t)}$ associated to the temporal modulation of the electric field magnitude, another term $\mathbf{f}_{(m)}$ associated with the nonlinear magnetic induced moment, and another term $\mathbf{f}_{MMP}$ associated with the spatial variation of the background magnetic field \citep{Espinoza-Troni_2023, Espinoza-Troni_etal_2024}, thus
\begin{equation}
    \mathbf{f} = \mathbf{f}_{(s)} + \mathbf{f}_{(t)} + \mathbf{f}_{(m)} + \mathbf{f}_{MMP}.
\label{eq:FPwashimikarpman}
\end{equation}
In this work, we neglect the PF temporal term of $\mathbf{f}_{(t)}$, since the wave amplitude is regarded as stationary. This approximation is considered for simplicity and is valid when assuming the wave amplitude temporal modulation is much slower than the transit time of the wave packet over a characteristic length of the field (as also done by \citet{Guglielmi_Hayashi_Lundin_Potapov_1999,Nekrasov_2012}). We also omit $\mathbf{f}_{(m)}$ since this term acts perpendicular to the background magnetic field, and does not contribute to the force balance equation along the geomagnetic field lines. The spatial term is given by
\begin{equation}
    \mathbf{f}_{(s)} = \frac{1}{16\pi}(\varepsilon_{ij} - \delta_{ij})\nabla \hat{E}^*_{i}\hat{E}_{j},    \label{eq:FPespacial}
\end{equation}
where $\varepsilon_{ij}$ are the components of the dielectric tensor and $\delta_{ij}$ is the Kronecker delta \citep{Washimi_Karpman_1976}. This contribution is well-known in nonlinear laser-plasma interactions, where it can induce density modifications \citep{Hora_1969}. Modifying the local dielectric tensor also affects wave propagation in plasmas. The remaining contribution $\mathbf{f}_{MMP}$ originates from the interaction between the background magnetic field gradient and the nonlinear magnetic moment induced in the particles by the wave. This term is usually called "Magnetic Moment Pumping" (MMP) \citep{Lundin_Guglielmi_2007}, and is given by
\begin{equation}
    \begin{split}
        \mathbf{f}_{(MMP)} &= M\nabla |\av{\mathbf{B}}|,
    \end{split}
\end{equation}
where $M = (1/16\pi)(\partial \varepsilon_{ij}/\partial |\av{\mathbf{B}}|) E^*_{i}E_j$ is the magnetic moment induced by the wave. For EMIC waves, this term points towards regions of decreasing magnetic field strength \citep{Espinoza-Troni_etal_2024}. For this reason, it has been proposed as part of a mechanism that accelerates ions in the polar regions, thereby contributing to the polar wind \citep{Lundin_Hultqvists_1989,Guglielmi_Lundin_2001,Guglielmi_Lundin_Potapov_2004}. It may also lead to plasma accumulation in the minimum of the geomagnetic field, which will be the equator for a dipolar magnetosphere, as discussed in \citet{Guglielmi_etal_1995,Guglielmi_Hayashi_Lundin_Potapov_1999,Nekrasov_Feygin_2016}, and analyzed below. 

\section{Ion cyclotron dispersion relation for Kappa distributed plasmas}\label{sec:4}

In this investigation, we will consider EMIC waves propagating along the geomagnetic field lines of different planets in a plasma characterized by an isotropic Kappa distribution,
\begin{equation}
    \mathcal{F}_{\kappa \alpha}(\mathbf{v}) = \frac{\av{n_s}}{\pi^{3/2}v_{th\alpha}^3 \kappa^{3/2}}\frac{\Gamma(\kappa+1)}{\Gamma(\kappa-1/2)}\left(1 + \frac{v^2}{\kappa v_{th\alpha}^2}\right)^{-(\kappa + 1)} 
\label{eq:DistribucionKappa}
\end{equation}
Here, $\mathcal{F}_{\kappa \alpha}$ is the Kappa distribution for the species $\alpha$, $v_{th\alpha}\{= \sqrt{2k_B T_\alpha /m_\alpha}\}$ is their thermal velocity, $T_\alpha$ is their temperature and $\Gamma$ is the Gamma function. This is the simplest form of the family of Kappa distributions, with the Maxwellian distribution recovered in the limit $\kappa \rightarrow \infty$, and is valid for $\kappa > 3/2$. Planetary magnetospheres are often well modeled by combinations or modifications of Kappa distributions \citep{lazar_kappa_2021}. Although real distributions may exhibit temperature anisotropies and depend strongly on the specific region and space-weather conditions, here we adopt this isotropic form as a first step to assess the effects of suprathermal populations found in planetary magnetospheres on ponderomotive interactions. 

\citet{Espinoza-Troni_etal_2024} derived the dielectric tensor for Kappa distributed plasmas under the low-temperature approximation $kv_{th\alpha}/(\omega \pm |\Omega_{\alpha}|) \ll 1 $, where $\Omega_{\alpha} = q_\alpha |\av{\mathbf{B}}|/m_\alpha c$ is the gyrofrequency of the species $\alpha$. For a left-handed polarized wave, with the carrier wave frequency lower than the ion gyrofrequency in a proton-electron plasma, the dielectric eigenvalue of EMIC waves with field-aligned propagation is approximated as
\begin{equation}
\varepsilon(\omega,k) \approx \varepsilon_c(\omega) - \frac{k^2c^2}{\omega^2}\delta_\kappa(\omega),
\label{eq:varepsilon}
\end{equation}
where $\varepsilon_c(\omega)$ is the EMIC dielectric eigenvalue for cold plasmas is 
\begin{equation}
\varepsilon_c(\omega) = 1 + \left(\frac{c}{c_A}\right)^2\frac{\Omega_i}{(\Omega_i-\omega)},
\label{eq:varepsilon_cold_EMIC}
\end{equation}
and $\delta_\kappa(\omega)$ is the finite temperature correction with the non-thermal effects
\begin{equation}
    \delta_\kappa(\omega) = -\left(\frac{\kappa}{\kappa-3/2}\right)\frac{\beta}{2}\frac{\omega \Omega_i^2}{(\Omega_i-\omega)^3},
\label{eq:delta_finite_temperature_EMIC}
\end{equation}
with $\beta = 8\pi n k_B T/|\av{\mathbf{B}}|^2$ the plasma beta. On the other hand, the dispersion relation is given by
\begin{equation}
\varepsilon(\omega,k) = \frac{k^2c^2}{\omega^2}.
\end{equation}
Using this expression, we can include the dispersion relation in Equation \eqref{eq:varepsilon}, and since our approximation is consistent for low-beta regimes, we can consider up to the second order of $\delta_\kappa(\omega)$. Then, we have that
\begin{equation}
\varepsilon(\omega,k(\omega)) = \frac{\varepsilon_c}{(1+\delta_\kappa)} \approx \varepsilon_c\left(1-\delta_\kappa\right).
\label{eq:varepsilon_approx}
\end{equation}
We will also introduce the following normalization. We will denote $\mathcal{B} = |\av{\mathbf{B}}|/B_0$ as the normalized background magnetic field, $\Omega_{i0} = qB_0/mc$ the ion gyrofrequency at the equator and $\bar{\omega} = \omega/\Omega_{i0} < 1$ the normalized frequency with respect to the equator gyrofrequency. Therefore, from Equations \eqref{eq:varepsilon_cold_EMIC} and \eqref{eq:delta_finite_temperature_EMIC} we have that
\begin{equation}    \varepsilon_c(\omega) \approx  \left(\frac{c}{c_{A0}}\right)^2 \frac{\bar{n}}{\mathcal{B}(\mathcal{B}-\bar{\omega})} = \left(\frac{c}{c_{A0}}\right)^2K\bar{n},
\label{eq:varepsc_eq}
\end{equation}

\begin{equation}
    \delta_\kappa(\omega) = -\frac{\beta_{\kappa 0}}{2}\frac{\bar{\omega}\bar{n}}{\left(\mathcal{B}-\bar{\omega}\right)^3} = \beta_{0 \kappa}P\bar{n},
\label{eq:deltak_eq}
\end{equation}
where we have considered that $c/c_{A0} \gg 1$ and we have defined $K = 1/[\mathcal{B}(\mathcal{B}-\bar{\omega})]$ and $P = -(1/2)\bar{\omega}/(\mathcal{B}-\bar{\omega})^3$. These expressions are valid away from the resonance region $\mathcal{B} \approx \bar{\omega}$, where the approximation breaks down; our PF expression implicitly assumes an adiabatic approximation, which is violated near gyro-resonance \citep{Lamb_etal_1984,Kono_and_Sanuki_1987}. 

\section{Ponderomotive force for ion cyclotron waves}\label{sec:5}

In this section, we calculate the various PF terms for EMIC waves in a low-beta, non-thermal plasma. \citet{Espinoza-Troni_etal_2024} computed the coefficients accompanying the spatial and temporal modulation of the wave in the PF for an isotropic Kappa plasma. Nevertheless, to obtain the full expression for the PF and complete the set of equations, we need the magnitude of the wave's electric field $\mathbf{\hat{E}}(\mathbf{r})$ as a function of the background variables. For obtaining such an expression, we can notice that we are assuming that $\mathbf{\hat{E}}(\mathbf{r})$ has a left-handed polarization
\begin{equation}
    \mathbf{\hat{E}}(\mathbf{r}) = \hat{E}(x_\parallel)\left(\mathbf{\hat{x}}_{\perp 1} - i\mathbf{\hat{x}}_{\perp 2} \right),
\end{equation}
where $\mathbf{\hat{x}}_{\perp 1}$ and $\mathbf{\hat{x}}_{\perp 2}$ are orthogonal coordinates aligned perpendicular to the magnetic field, with $x_\parallel$ the distance along the magnetic field lines. Also, we will consider that $\mathbf{\hat{J}} = \frac{\omega}{4\pi i}\left[\varepsilon(\omega,k)-1\right]\mathbf{\hat{E}}$. 
To obtain this last relation, one must neglect the spatial and temporal variations of the slow-time-scale quantities when Fourier-transforming the first-order perturbations within the kinetic framework of the Vlasov equation \citep{BOOK_krall_trivelpiece_1986}. Including these variations would require a more sophisticated analysis, analogous to the approach of \citet{Lee_Nam_Parks_1998}, in which the wave modulation is governed by a generalization of the derivative nonlinear Schrödinger equation (DNLS). However, in our case, such a procedure would need to be extended using a full kinetic formalism, which is beyond the scope of this work. Therefore, using Equation \eqref{eq:WaveEq} and neglecting the curvature of the geomagnetic field lines, we obtain
\begin{equation}
    \frac{\partial^2 \hat{E}}{\partial x_\parallel^2} + k^2(x_\parallel)\hat{E} = 0.
\end{equation}
As in other studies of plasma density redistribution by the PF, for simplicity purposes, we neglect the curvature of the geomagnetic field \citep{Nekrasov_2012}. This is valid if the curvature of the geomagnetic field lines varies very slowly compared to the wave modulation, and is useful as a first-order approach for studying finite-temperature and non-thermal effects of the PF on the magnetospheric density redistribution of some planets due to traveling waves. We can notice that in this case, $\varepsilon$ (and therefore $k^2$) depends on slow-time scale variables, which should vary slowly in a wavelength for the Washimi-Karpman formalism to be valid. Therefore, we can assume that $(1/k)(d\varepsilon/dx_\parallel) \ll \varepsilon$, which is the condition for the WKB approximation \citep{Ghatak_Gallawa_Goyal_1991}. Then, using this approximation, we can show that
\begin{equation}
    |\hat{E}(x_\parallel)| \propto \frac{1}{k^{1/2}(x_\parallel)} \propto \frac{1}{\varepsilon^{1/4}(x_\parallel)}.
\end{equation}
Also, the magnetic field wave amplitude $|\hat{B}|$ is related to the electric field wave amplitude by
\begin{equation}
    |\hat{E}| = \frac{|\hat{B}|}{\varepsilon^{1/2}}.
\end{equation}
Therefore we can express $|\hat{E}|^2$ as
\begin{equation}
    |\hat{E}|^2 = \left(\frac{B_0^2 \nu^2}{\varepsilon^{1/2}(x_\parallel = 0)}\right) \frac{\alpha}{\varepsilon^{1/2}},
\end{equation}
where $\nu = |\hat{B}(x_\parallel = 0)|/B_0 \sim \zeta$ is the ratio of the magnetic field wave amplitude to the background equatorial magnetic field. In our approach, we set $\alpha=1$ (as in \citep{Nekrasov_2012}); however, we retain the explicit term $\alpha$ because \citet{Lundin_Guglielmi_2007} derive $\alpha = \mathcal{B}$ by assuming a vanishing divergence for the low-time-scale Poynting flux. Notably, both approaches yield qualitatively consistent results (not shown). Since the PF is a second-order effect, this parameter should be small. Neglecting the terms of the order of $\delta_\kappa^2$ since we are in a low beta plasma regime, we can approximate $\varepsilon^{-1/2} \approx \varepsilon_c^{-1/2}\left(1+\delta_\kappa/2\right)$, then
\begin{equation}
    |\hat{E}|^2 \approx B_0^2\nu^2\left(\frac{c}{c_{A0}}\right)^{-1}\frac{(1-\bar{\omega})^{1/2}}{\varepsilon_c^{1/2}}\left[\left(1+\frac{\delta_{\kappa 0}}{2}\right)+\frac{\delta_\kappa}{2}\right]\alpha,
\label{eq:E_modulation}
\end{equation}
where $\delta_{\kappa 0} = \delta_\kappa(x_\parallel = 0) = -(1/2)\beta_{\kappa 0}\bar{\omega}/(1-\bar{\omega})^3$.

\subsection{Spatial term}

Notice that the fast-time scale electric field that we are considering is an eigenvector of the dielectric tensor $\varepsilon_{ij}$, then the spatial term of the PF will be given by \citep{Espinoza-Troni_etal_2024}
\begin{equation}
    f_{(s)\parallel} = \frac{1}{16\pi}(\varepsilon-1)\nabla_\parallel |\hat{E}|^2.
\end{equation}
Using equations \eqref{eq:E_modulation} and \eqref{eq:varepsilon_approx}, and neglecting $\delta_\kappa^2$ we have that
\begin{equation}
\begin{split}
    \bar{f}_{(s)\parallel} &= \left(A_1+\beta_{\kappa 0} A_2\right)\bar{n}^{1/2} + \beta_{\kappa 0}A_3\bar{n}^{3/2} \\
    &+\left(A_4+\beta_{\kappa 0}A_5\right)\bar{n}^{-1/2}\bar{\nabla}_\parallel \bar{n} + \beta_{\kappa 0}A_6\bar{n}^{1/2}\bar{\nabla}_\parallel \bar{n},
\label{eq:F_s}
\end{split}
\end{equation}
where the functions $A_i(x_\parallel)$ (for $i=1,...,6$) are defined in Appendix \ref{sec:AppendixA} with more details of the calculation. The functions $A_i(x_\parallel)$ depend on the the normalized background magnetic field magnitude $\mathcal{B}(x_\parallel)$ and the normalized frequency $\bar{\omega}$.

\subsection{Magnetic moment pumping}

The MMP term of the PF is given by \citep{Espinoza-Troni_etal_2024}
\begin{equation}
    f_{(MMP)\parallel} = \frac{|\hat{E}|^2}{16\pi}\frac{\partial \varepsilon}{\partial \mathcal{B}}\bar{\nabla}_\parallel \mathcal{B},
\label{eq:MMP_force_term}
\end{equation}
then, replacing the expression for the wave modulation \eqref{eq:E_modulation} and the finite temperature dielectric tensor eigenvalue \eqref{eq:varepsilon_approx}, we have that
\begin{equation}
    f_{(MMP)\parallel} = \left(A_7 + \beta_{\kappa 0}A_8\right)\bar{n}^{1/2} + \beta_{\kappa 0}A_9\bar{n}^{3/2},
\label{eq:F_MMP}
\end{equation}
where functions $A_7(x_\parallel)$, $A_8(x_\parallel)$, and $A_9(x_\parallel)$ are defined in Appendix \ref{sec:AppendixA} and also depend on the normalized background magnetic field magnitude $\mathcal{B}(x_\parallel)$ and the normalized frequency $\bar{\omega}$.

\subsection{Total ponderomotive force}

Summing expressions \eqref{eq:F_s} and \eqref{eq:F_MMP} for the spatial and MMP PF terms, respectively, we obtain that the total PF along the geomagnetic field lines $f_\parallel = f_{(s)\parallel} + f_{(MMP)\parallel}$ is given by
\begin{equation}
\begin{split}
    &f_\parallel = \left[(A_1+A_7)+\beta_{\kappa 0}(A_2+A_8)\right]\bar{n}^{1/2} + \beta_{\kappa 0}\left(A_3+A_9\right)\bar{n}^{3/2} \\
    &+\left(A_4+\beta_{\kappa 0}A_5\right)\bar{n}^{-1/2}\bar{\nabla}_\parallel \bar{n}+\beta_{\kappa 0}A_6\bar{n}^{1/2}\bar{\nabla}_\parallel \bar{n}.
\label{eq:Total_PF}
\end{split}
\end{equation}

\section{Plasma density redistribution for non-thermal plasmas}\label{sec:6}

\subsection{Differential equation for density redistribution}

In this section, we derive and analyze the differential equation governing the redistribution of plasma density along geomagnetic field lines in the magnetospheres of some planets. We then discuss some analytical properties of this equation.  

Using the force balance equation (Eq. \ref{eq:Force_balance_normalized}) and the expression for the total PF (Eq. \ref{eq:Total_PF}), we obtain the following differential equation for the slow-time-scale plasma density redistribution along the geomagnetic field lines
\begin{equation}
    \frac{d\bar{n}}{dx_\parallel} = \frac{C_g \bar{g}_\parallel (x_\parallel) \bar{n} + \nu^2\Phi_1(x_\parallel,\beta_{\kappa 0})\bar{n}^{1/2}+\nu^2\Phi_2(x_\parallel,\beta_{\kappa 0})\bar{n}^{3/2}}{\beta_{\kappa 0}+\nu^2\Phi_3(x_\parallel,\beta_{\kappa 0})\bar{n}^{-1/2}+\nu^2 \Phi_4(x_\parallel,\beta_{\kappa 0}) \bar{n}^{1/2}},
\label{eq:ODE}
\end{equation}
where the functions $\Phi_i(x_\parallel,\beta_{\kappa0})$ are given by
\begin{equation}
    \Phi_1(x_\parallel,\beta_{\kappa 0}) = \left(A_1(x_\parallel)+A_7(x_\parallel)\right) + \beta_{i 0\kappa}\left(A_2(x_\parallel)+A_8(x_\parallel)\right),
\end{equation}

\begin{equation}
    \Phi_2(x_\parallel,\beta_{\kappa 0}) = \beta_{\kappa 0}\left(A_3(x_\parallel)+A_9(x_\parallel)\right),
\end{equation}

\begin{equation}
    \Phi_3(x_\parallel,\beta_{\kappa 0}) = -\left(A_4(x_\parallel)+\beta_{\kappa 0}A_5\right), 
\end{equation}

\begin{equation}
    \Phi_4(x_\parallel,\beta_{\kappa 0}) = -\beta_{\kappa 0} A_6(x_\parallel).
\end{equation}
Notice that in the cold plasma limit ($\beta_{\kappa 0} \rightarrow 0$), our results differ from those of \citet{Guglielmi_Hayashi_Lundin_Potapov_1999}. While their approach restricts thermal effects solely to the pressure term, our general model incorporates the plasma beta into both the pressure and the ponderomotive force. Therefore, taking $\beta_{\kappa 0} \rightarrow 0$ in our framework removes both the macroscopic fluid pressure and the thermal corrections to the ponderomotive force. This reduces the system to a strict equilibrium between the gravity force and the cold ponderomotive force. Then, for a cold plasma, we obtain that
\begin{equation}
\frac{d\bar{n}}{dx_\parallel} = \left(\frac{C_g}{\nu^2}\right)\frac{\bar{g}_\parallel(x_\parallel)}{\Phi_3(x_\parallel,0)} \bar{n}^{3/2} +  \frac{\Phi_1(x_\parallel,0)}{\Phi_3(x_\parallel,0)}\bar{n}.    
\end{equation}
When gravity is weak compared to the PF, i.e, $(C_g/\nu^2) (\bar{g}_\parallel \bar{n}^{1/2}/\Phi_1)\ll 1$, we still can obtain an equilibrium for the density given by
\begin{equation}
    \frac{d\bar{n}}{dx_\parallel} = \frac{\Phi_1(x_\parallel,0)}{\Phi_3(x_\parallel,0)}\bar{n},
\end{equation}
Therefore, according to our differential equation, in a cold plasma in equilibrium along the geomagnetic field lines with a strong PF compared to the effect of gravity, we will have that the density satisfies the following relation
\begin{equation}
    \bar{n} = \exp\left(\int_0^{x_\parallel}\frac{\Phi_1(s,0)}{\Phi_3(s,0)}ds \right).
\label{eq:Cold_Withoutg_case}
\end{equation}
In that case, there is an equilibrium between the spatial term of the PF and the MMP term. This solution resembles the form of the non-PF solution, but instead of having the gravity acceleration inside the integral, it has $\Phi_1(x_\parallel,0)/\Phi_3(x_\parallel,0)$ as we can see comparing Equation \eqref{eq:Cold_Withoutg_case}  with the $\nu = 0$ solution
\begin{equation}
    \bar{n} = \exp\left(\frac{C_g}{\beta_{\kappa 0}}\int_0^{x_\parallel} \bar{g}_\parallel (s) ds\right).
\label{eq:Solution_v=0}
\end{equation}
We now analyze the ODE's behavior with respect to its parameters. Following the procedure in \citet{Guglielmi_Hayashi_Lundin_Potapov_1999}, it is useful to study the nullclines to determine the extreme values of the density and whether they correspond to minima or maxima. Considering $dn/dx_\parallel = 0$ in Eq.~\eqref{eq:ODE}, we obtain that the nullcline $n_c(x_\parallel)$ ( i.e, the possible values for extrema of the solution) is given by
\begin{equation}
    n_c(x_\parallel) = \left(\frac{C_g}{\nu^2}\right)^2\frac{\bar{g}_\parallel^2}{4\Phi_2^2}\left\{\sqrt{1-\left(\frac{C_g}{\nu^2}\right)^{-2}\frac{4\Phi_1\Phi_2}{\bar{g}_\parallel^2}} - 1\right\}^2.
\label{eq:nullcline}
\end{equation}
In case there is a minimum of the geomagnetic field at the equator, as in a dipolar model of the magnetosphere, the ODE for the density will also have a nullcline at $x_\parallel = 0$. The character of the solution will depend on whether the nullcline is above or below the initial condition $\bar{n} = 1$ for $x_\parallel = 0$, and on the sign of $d\bar{n}/dx_\parallel$ below and above the nullcline, as well as the characteristics of its slope. Therefore, it will be useful to calculate the critical value of $\Lambda = \nu^2/C_g$  as a function of $\bar{\omega}$ and $\beta_{\kappa 0}$ for which the nullcline at $x_\parallel = 0$ coincides with the initial condition, i.e $n_c(x_\parallel = 0) = \bar{n}(x_\parallel=0)=1$. Notice that $\Lambda$ quantifies the relative effect of gravity compared to the PF in the density redistribution. To perform a specific analysis of the nullclines, we first need to specify a particular geomagnetic field model. First, we define
\begin{equation}
    \xi_1 =  \lim_{x_\parallel\rightarrow 0}\frac{\bar{g}_\parallel}{2\Phi_2}, 
\end{equation}

\begin{equation}
    \xi_2 = \lim_{x_\parallel \rightarrow 0} \frac{\bar{g}_\parallel}{2\sqrt{-\Phi_1 \Phi_2}}\,.
\end{equation}

Then we obtain that the critical value $\Lambda_c$ that has to reach $\Lambda$ for the nullcline to match the initial condition at $x_\parallel = 0$ is given by
\begin{equation}
    \Lambda_c (\bar{\omega},\beta_{\kappa 0}) = \frac{2\xi_1}{(\xi_1/\xi_2)^2-1}\,,
\label{eq:epsilonc_critical_parameter}
\end{equation}
from which it is straightforward to calculate that 

\begin{equation}
    \xi_1(\bar{\omega},\beta_{\kappa 0}) = \frac{1}{\beta_{\kappa 0}}\left[\frac{4(1-\bar{\omega})^4}{\bar{\omega}(1/2-\bar{\omega})}\right]\left(\frac{\left.\bar{\nabla}_\parallel\bar{g}_\parallel\right|_{x_\parallel = 0}}{\left.\bar{\nabla}_\parallel^2\mathcal{B}\right|_{x_\parallel = 0}}\right)\,,
\end{equation}
and
\begin{equation}
\begin{split}
    &\xi_2(\bar{\omega},\beta_{\kappa 0}) = \left[\beta_{\kappa 0}\left(1-\beta_{\kappa 0}\frac{\bar{\omega}}{4(1-\bar{\omega})^3}\right)\right]^{-1/2} \\
    &\times \left[\frac{4(1-\bar{\omega})^{5/2}}{\bar{\omega}^{1/2}(2-\bar{\omega})^{1/2}(1/2-\bar{\omega})^{1/2}}\right]\left(\frac{\left.\bar{\nabla}_\parallel\bar{g}_\parallel\right|_{x_\parallel=0}}{\left.\bar{\nabla}_\parallel^2\mathcal{B}\right|_{x_\parallel = 0}}\right).
\end{split}
\end{equation}

\subsection{Dipolar model for the geomagnetic field}

Up to this point, we have not assumed any particular geomagnetic field model. As a first approach, we consider the simple case of a dipolar magnetic field centered at the planet and directed along $\hat{\mathbf{z}}$. Then,
\begin{equation}
    \mathcal{B}_r = 2L^3\left(\frac{1} {\bar{r}}\right)^3\cos\theta,
\end{equation}

\begin{equation}
    \mathcal{B}_\theta = L^3 \left(\frac{1}{\bar{r}}\right)^3 \sin\theta,
\end{equation}
and
\begin{equation}
    \mathcal{B} = L^3\left(\frac{1}{\bar{r}}\right)^3\sqrt{1+3\cos^2\theta},
\end{equation}
with $\mathcal{B}_r =\mathbf{\hat{r}}\cdot \av{\mathbf{B}}$ and $\mathcal{B}_\theta =\mathbf{\hat{\theta}}\cdot \av{\mathbf{B}}$, where $\bar{r} = r/R_p = L\sin^2\theta$ is the normalized radius with respect to the planet radius, $\theta$ is the polar angle, and $L$ is the L-shell parameter \citep{McIlwain_1961}. Then, $\bar{r}(x_\parallel = 0) = L$ and $\theta(x_\parallel = 0) = \pi/2$. Therefore, considering the above expressions and that $\mathbf{g} = GM_p/r^2 \hat{\mathbf{r}}$, we can obtain $\bar{g}_\parallel$ and $\mathcal{B}$ as functions of the polar angle along a certain magnetic field line defined by the parameter $L$, with   
\begin{equation}
    \bar{g}_\parallel(\theta) = -\frac{1}{L^2\sin^4\theta}\frac{2\cos\theta}{\sqrt{1+3\cos^2\theta}},
\end{equation}

\begin{equation}
    \mathcal{B}(\theta) = \frac{\sqrt{1+3\cos^2\theta}}{\sin^6\theta}.
\end{equation}

\begin{equation}
    \frac{d}{dx_\parallel} = \frac{d\theta}{dx_\parallel}\frac{d}{d\theta} = \frac{1}{L \sin\theta \sqrt{1+3\cos^2\theta}}\frac{d}{d\theta}.
\end{equation}
We can solve the differential equation (Eq. \ref{eq:ODE}) to obtain the plasma density redistribution along the geomagnetic field lines using these expressions.

\begin{figure}[ht]
\centering
\includegraphics[width=0.8\linewidth]{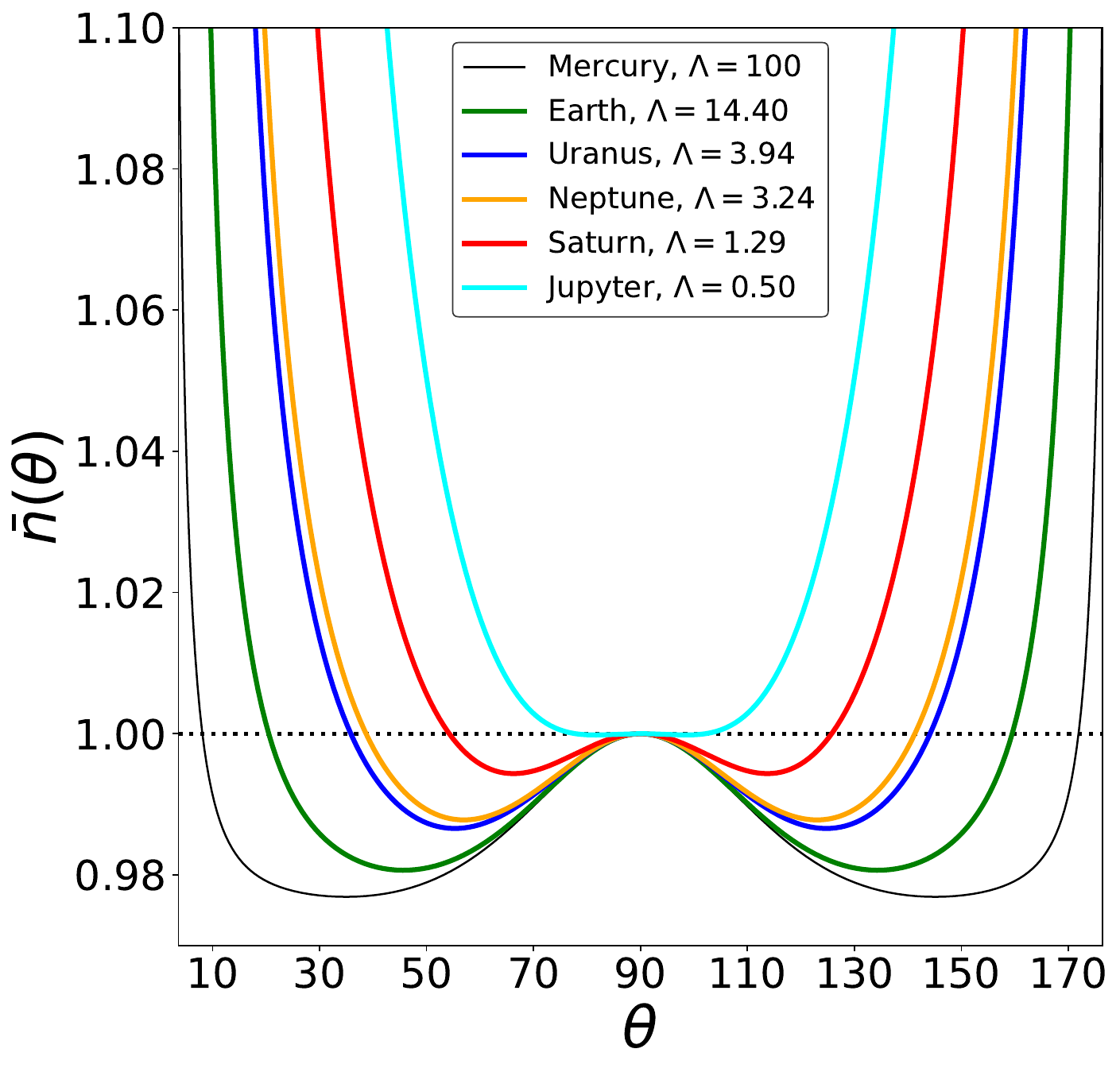}
    \caption{Normalized density redistribution along the geomagnetic field lines as a function of the latitude for different values of $\Lambda$ depending on the planet for $\nu=0.1$, $\bar{\omega} = 0.1$, $L=2$, $c/c_{A0} = 10^3$, $\beta_{0} = 0.1$ and a Maxwellian distribution.}
\label{fig:Plot_n_planets}
\end{figure}

\begin{figure*}[ht]
\centering
\includegraphics[width=0.7\linewidth]{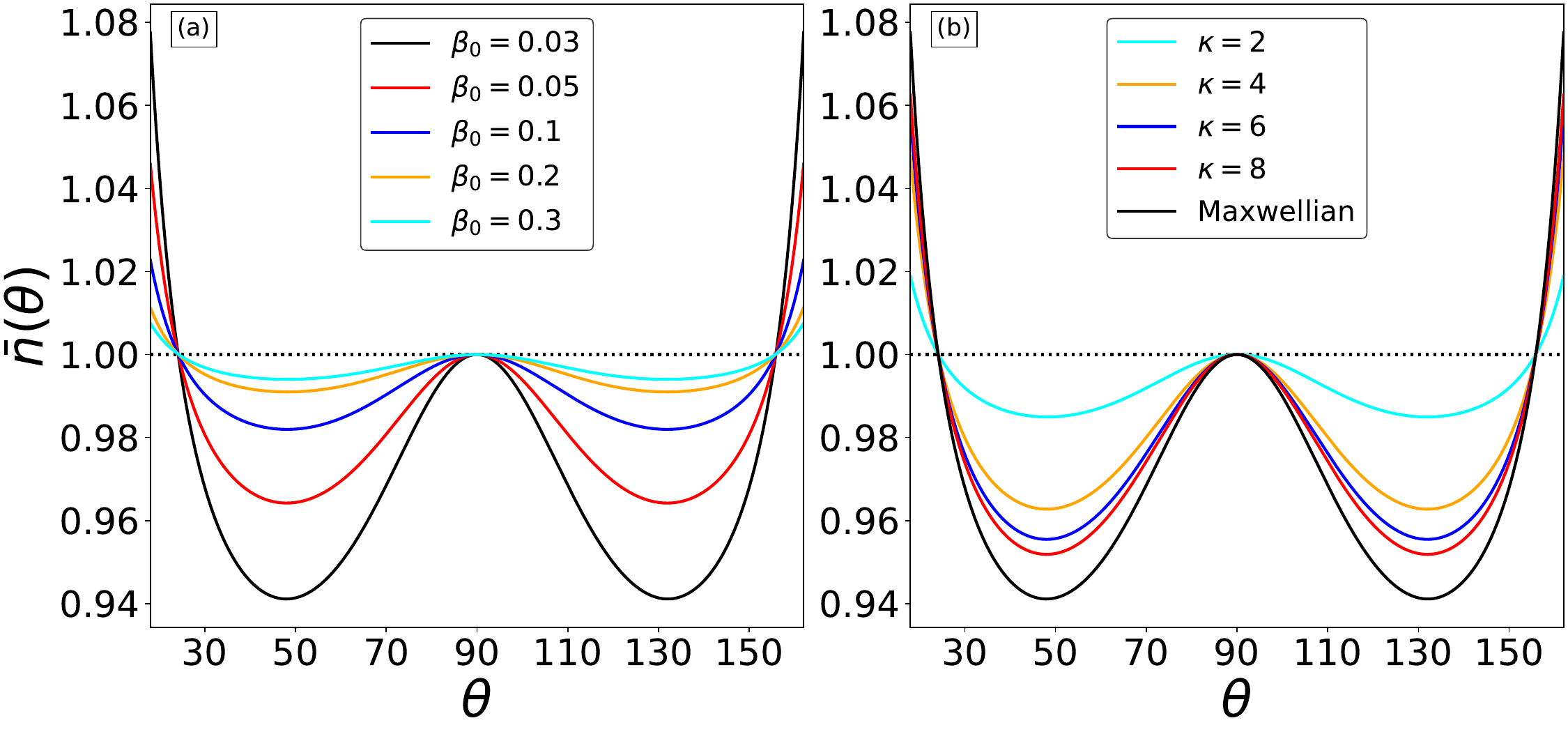}
    \caption{Normalized density redistribution along the geomagnetic field lines as function of the colatitude for $\nu=0.1$, $\bar{\omega} = 0.1$, $L=2$, $c/c_{A0} = 10^3$ and for (a) different values of $\beta_{0}$ with a Maxwellian distribution (b) different values of $\kappa$, with $\beta_{0} = 0.03$.}
\label{fig:Plot_n}
\end{figure*}

\begin{figure}[ht]
\centering
\includegraphics[width=0.7\linewidth]{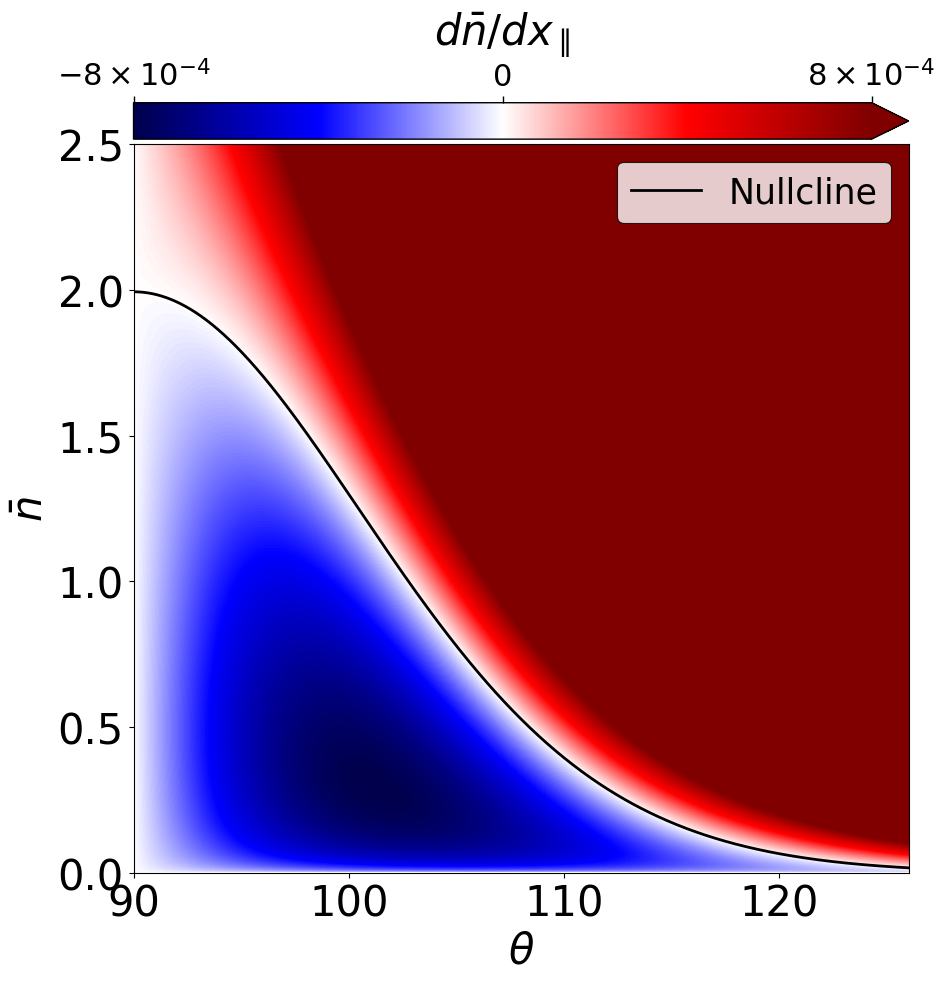}
\caption{Shaded isocontours of $d\bar{n}/d\bar{x}_\parallel$ as a function of the colatitude and the normalized density with the nullcline $n_c$ represented with the black curve; with $\nu = 0.1$, $\bar{\omega} = 0.1$, $c/c_{A0} = 10^3$, $\beta_{i0\kappa} = 0.1$, $L = 2$ and the $C_g$ of Earth.}
\label{fig:Plot_nullclines_colormap}
\end{figure}

\begin{figure*}[ht]
\centering
\includegraphics[width=0.8\linewidth]{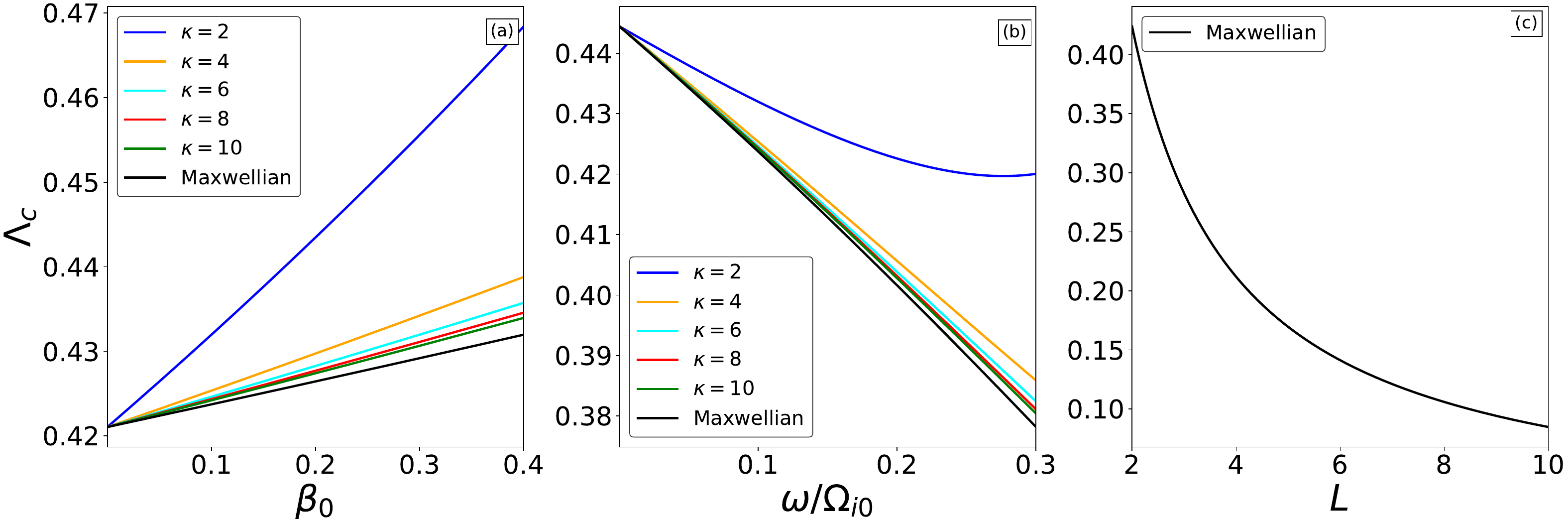}
\caption{Critical value $\Lambda_c$ for different values of the kappa parameter as a function of: (a) the plasma beta $\beta_0$, with $\bar{\omega} = 0.1$ and $L=2$; (b) the normalized frequency $\omega/\Omega_{i0}$, with $\beta_{0} = 0.1$ and $L = 2$; and (c) the L-shell parameter $L$, with $\beta_{0} = 0.1$ and $\bar{\omega} = 0.1$. In panel (c), only the Maxwellian case is displayed, as the curves for other $\kappa$ values overlap and are omitted for clarity.}
\label{fig:Plot_epsilonc}
\end{figure*}

\subsection{Numerical solutions}

In this section, we analyze the numerical solutions of the differential equation (Eq. \ref{eq:ODE}) and their behavior for different values of $\beta_{\kappa 0}$, $L$, and $\bar{\omega}$. To solve the ODE describing plasma density redistribution along geomagnetic field lines, we use a fourth-order Runge-Kutta method \citep{Butcher_1996} in the presence of a dipolar magnetic field. Since the dipolar geomagnetic field model is more appropriate for the plasmasphere, we must consider low $L$ values. 

\renewcommand{\tabcolsep}{10pt}
\begin{table}
\centering
\caption{Parameters used to calculate $C_g$ for each planet.}
\label{tab:PlanetsParameters}
\begin{tabular}{cccc}
\toprule
Planet & $R_p$[m] & $M_p$[kg] & $C_g$ \\ 
\midrule
Mercury & $2.44\cdot10^{6}$ & $3.30\cdot10^{23}$ & $1.00\cdot 10^{-4}$ \\
Earth & $6.38\cdot10^{6}$ & $5.97\cdot10^{24}$ & $6.94\cdot 10^{-4}$ \\
Jupiter & $6.99\cdot10^{7}$ & $1.90\cdot10^{27}$ & $2.01\cdot 10^{-2}$ \\
Saturn & $5.44\cdot10^{7}$ & $5.68\cdot10^{26}$ & $7.75\cdot 10^{-3}$ \\
Uranus & $2.54\cdot10^{7}$ & $8.68\cdot10^{25}$ & $2.54\cdot 10^{-3}$\\
Neptune & $2.46\cdot10^{7}$ & $1.02\cdot 10^{26}$ & $3.08\cdot 10^{-3}$ \\
\bottomrule
\end{tabular}
\tablefoot{We consider $c/c_{A0} = 10^{3}$, where the planetary radius $R_p$ is in meters [m], and the planetary mass $M_p$ in kilograms [kg].}
\end{table}	

Figure \ref{fig:Plot_n_planets} shows the plasma density redistribution as a function of colatitude for $\nu = 0.1$, and different values of $C_g$, depending on the mass and radius of each planet in the solar system with a magnetosphere. We use $\nu = 0.1$ as an upper limit of typical observed wave amplitudes to illustrate this PF effect. A detailed study on $\nu$ dependence will be given in section \ref{sec:7}. The parameters for each planet are listed in Table \ref{tab:PlanetsParameters}. In this figure, we use the same plasma parameters for all planets to highlight the effect of gravity relative to the PF due to suprathermal particles, rather than focusing on the specific characteristics of each planetary magnetosphere. This approach allows us to maintain a certain generality in the discussion. We consider low-temperature plasmas with $\beta_{\kappa 0} = 0.1$, and Alfvén velocity of $c/c_{A0} = 10^3$, and a low L-shell parameter $L = 2$.  Although these parameters do not necessarily represent typical magnetospheric values for each region of each planet, they can be applied across a wide range of zones in different planetary magnetospheres. For instance, in Jupiter, the inner magnetosphere extends to $L\sim 10$, with plasma beta mostly below $\sim 0.2$ due to the strong background magnetic field and low temperatures of the torus plasma region produced by the Io moon \citep{Khurana_etal_2004}. Nevertheless, both ion and electron plasma beta increase with $L$ in the inner Jovian magnetosphere, ranging from $\beta \sim  0.01$ to $\beta \sim 1$ \citep{Mauk_etal_1996}. For Saturn, data from Pioneer 11 and Voyagers 1 and 2 indicate that the inner magnetosphere extends to $L\sim 4$, with cold plasma characterized by $\beta < 1$ \citep{Schardt_1983}. The icy giants, Uranus and Neptune, exhibit similar magnetospheric properties, both being at large heliocentric distances and occupying different regions of parameter space compared to the inner planets in the solar system \citep{Arridge_etal_2021}. The low-beta approximation is particularly valid in their magnetospheres, as both are very tenuous, with plasma beta values lower than those of Jupiter and Saturn. Indeed, Voyager 2 measured within the magnetospheres of Uranus and Neptune a plasma beta at most of $\beta \sim 0.1$ (for $L<15$) and $\beta \sim 0.2$ respectively \citep{Krimigis_etal_1986, Krimigis_etal_1989}. 

Figure \ref{fig:Plot_n_planets} shows that plasma density, instead of reaching a minimum at the equator (as expected in the absence of the PF), exhibits a local maximum. This indicates that the PF produces an equatorial plasma concentration that decreases with colatitude, reaches a minimum, and then increases again as it approaches the planet's surface. These solutions have the same qualitative shape as those obtained with the cold PF studied in \citet{Guglielmi_Hayashi_Lundin_Potapov_1999}; therefore, the finite-temperature correction to the PF does not change the qualitative behavior of the solutions predicted in previous works. The density minimum is more pronounced for low-mass-density planets like Mercury or Earth (i.e, for larger $\Lambda$). For instance, in Mercury's dipolar magnetosphere approximation with $L=2$, $\beta_{0} =0.1$, $\omega/\Omega_{i0} = 0.1$, $\nu=0.1$ and $c/c_{A0} = 10^3$, latitudes of $\sim 60$ degrees exhibit a $\sim 2\%$ decrease of density relative to the equator. In these high-$\Lambda$ cases, the PF dominates gravity near the equator, whereas for denser planets such as Jupiter, gravity dominates and the density minimum has a higher value. When $\Lambda$ is sufficiently small, the PF is no longer able to concentrate plasma at the equator, and density exhibits a minimum at the equator with monotonic increase toward higher colatitudes, resembling qualitatively the standard solution without the PF effect ($\nu = 0$) given in Equation \eqref{eq:Solution_v=0}.

\begin{figure*}[ht]
\centering
\includegraphics[width=0.9\linewidth]{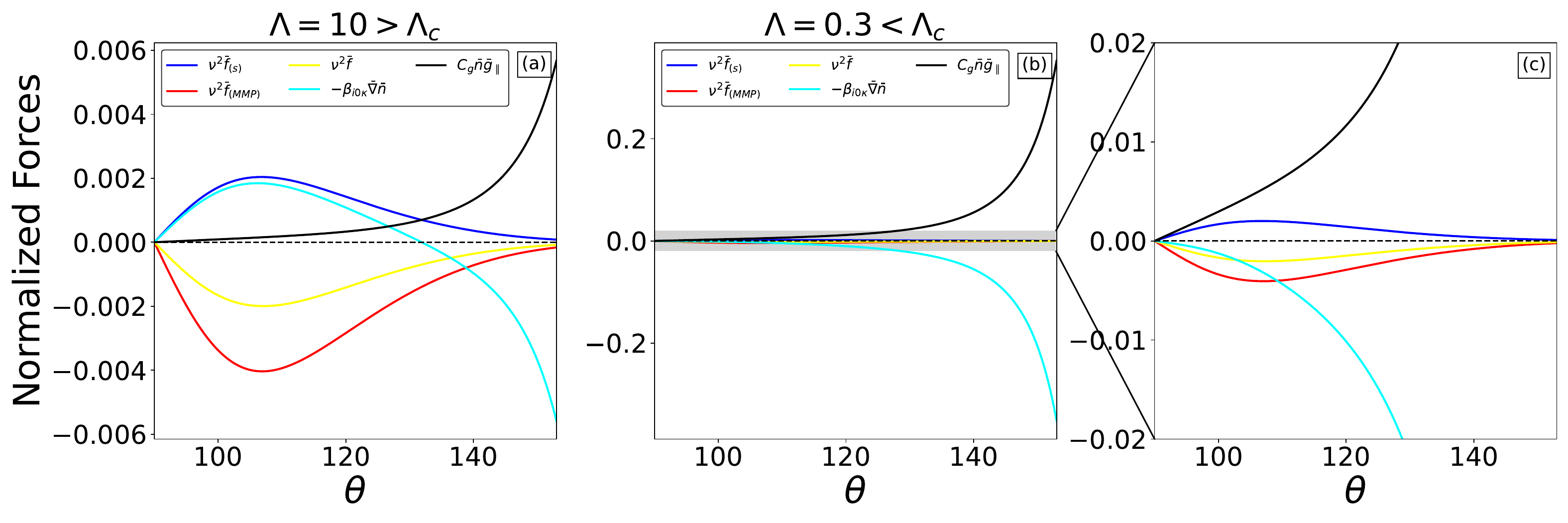}
    \caption{Panels (a)-(b): Normalized forces that act on the equilibrium Equation \eqref{eq:Force_balance_normalized}, as a function of the colatitude $\theta$ with $\nu=0.1$, $\bar{\omega} = 0.1$, $c/c_{A0} = 10^3$, $\beta_{i0\kappa} = 0.1$, $L=2$ for (a) $\Lambda = 10 >\Lambda_c$ (b) $\Lambda = 0.3 <\Lambda_c$. Panel (c): Zoom over the gray-shaded region in panel (b).}
\label{fig:Plot_Forces}
\end{figure*}

Panels (a) and (b) of Figure \ref{fig:Plot_n} show the normalized plasma density redistribution as a function of the colatitude $\theta$. Panel (a) evaluates the density for different values of the plasma beta $\beta_{0}$ for a Maxwellian distribution with $L=2$, $c/c_{A0} = 10^3$, $\nu = 0.1$, and $\bar{\omega}=0.1$. This panel shows that the magnitude of the density minimum decreases with increasing plasma beta. Therefore, our model predicts that PF effects on density redistribution along geomagnetic field lines decrease as plasma temperature increases. A similar trend can be expected as the kappa parameter decreases. Indeed, the kappa parameter can significantly suppress the density accumulation at the equator, as shown in panel (b) of Figure \ref{fig:Plot_n}, where the plasma density is plotted as a function of colatitude for different values of $\kappa$ with $\beta_0 = 0.03$, $L=2$, $c/c_{A0} = 10^3$, $\nu = 0.1$ and $\bar{\omega}=0.1$. As shown in this panel, for $\kappa = 2$, the equatorial density increase is less than $2\%$ compared to the $6\%$ observed in the Maxwellian case. Hence, the PF effect is less pronounced for plasmas with suprathermal distributions. This behavior can be understood by noting that as plasma beta increases or the kappa parameter decreases, plasma pressure increases, counteracting the PF's tendency to concentrate plasma at the equator. Conversely, when plasma beta decreases, the PF can concentrate more plasma at the equator, producing steeper spatial density variations. For very low plasma beta, the slow-scale density no longer varies smoothly in space, and the WKB approximation is invalid.

These results highlight the importance of including non-thermal correction in the PF, given that measurements of velocity distributions in planetary magnetospheres are often well fitted by Kappa distributions to describe their suprathermal tails \citep{lazar_kappa_2021}. For example, Cassini measurements at Jupiter near the equator for radial distances with $L\sim 6-46$ \citep{Mauk_etal_1996} indicate that a modified Kappa distribution accurately represents ions. This distribution has two spectral indices, and at high energies, a power law is recovered with an effective spectral index equal to the sum of the two indices. Using this sum, the kappa parameter can be considered in the range $\sim 2 - 4$. In Saturn, ion spectra are well fitted by a Kappa distribution with $\kappa \sim 6-8$ at radial distances with $L \sim 6-20$ \citep{Schardt_1983,Dialynas_etal_2009}. Therefore, both Jupiter and Saturn's magnetospheres exhibit significant suprathermal properties, which, under our approximations, are predicted to highly reduce the plasma confinement near the equator. As shown in panel (b) of Figure \ref{fig:Plot_n}, lower values of $\kappa$ lead to a broader latitudinal distribution, effectively redistributing the plasma density away from the equatorial plane. In the case of the icy giants, Voyager 2 measurements reveal nonthermal ion and electron spectra at Uranus \citep{Krimigis_etal_1986,Mauk_1987}, well modeled as a Maxwellian core with high-energy power-law tails (with exponents between 3 and 10). For Neptune, Voyager 2 data \citep{Krimigis_etal_1989} indicate that Triton influences the outer magnetosphere: inside the satellite's orbit ($L\sim 14$), the proton spectrum is Maxwellian, whereas outside it presents nonthermal properties and is modeled with a power-law distribution (with exponent $\sim 4$). Despite being modeled with a power law, Uranus and Neptune's high-energy tails are consistent with Kappa distributions \citep{lazar_kappa_2021}. Therefore, we expect that nonthermal properties in the planetary magnetospheres of the solar system will counteract the tendency of the PF to accumulate plasma at the equator in the dipolar magnetic field approximation. For more complex, realistic magnetosphere geometries, particularly the non-dipolar field of Neptune and Uranus \citep{Arridge_etal_2021}, and non-dipolar regions of the terrestrial magnetosphere, plasma accumulation due to the PF is expected to occur at local minima of the geomagnetic field, consistent with \citet{Nekrasov_Feygin_2016} using the \citet{Antonova_Shabanskii_1968} magnetic field model of the Earth's dayside magnetosphere. That study found density enhancements at local geomagnetic minima. Afterward, we will justify this point in more detail. 

To further understand the behavior of the solutions, we analyze their nullclines. Thus, we will analyze the properties of $d\bar{n}/dx_\parallel$, with the particular case of the dipolar geomagnetic field model. Figure  \ref{fig:Plot_nullclines_colormap} shows shaded isocontours of $d\bar{n}/dx_\parallel$ as a function of colatitude and the normalized density, with the nullcline of Equation \eqref{eq:nullcline} represented with the black curve; with $\nu = 0.1$, $\bar{\omega} = 0.1$, $c/c_{A0} = 10^3$, $\beta_{i0\kappa} = 0.1$, $L = 2$ and the $C_g$ of Earth. We focus on $\theta > \pi/2$, due to symmetry about the equator. Above the nullcline, the spatial derivative of the density is positive $d\bar{n}/dx_\parallel > 0$; below it is negative $d\bar{n}/dx_\parallel < 0$, with the nullcline having a negative slope. Hence, if the nullcline is below the initial condition at $x_\parallel = 0$, density increases monotonically with colatitude, exhibiting a minimum at the equator. Conversely, if the nullcline lies above the initial condition, density decreases until it crosses the nullcline, then increases indefinitely. For $\Lambda > \Lambda_c$, with $\Lambda_c$ defined in Equation \eqref{eq:epsilonc_critical_parameter}, we have that $n_c(x_\parallel = 0) < 1$, and there will be a maximum at the equator and two symmetrical minima around it. On the other hand, for $\Lambda < \Lambda_c$, then $n_c(x_\parallel = 0) > 1$, and the plasma density will have only a minimum located at the equator. The critical value $\Lambda_c$ depends on the plasma beta, kappa, the frequency, and the L-shell parameter. Consequently, as in the cold PF case studied by \citet{Guglielmi_Hayashi_Lundin_Potapov_1999}, the density closely resembles a second-kind phase transition. 

Panels (a)-(c) of Figure \ref{fig:Plot_epsilonc} show the critical value $\Lambda_c$ as a function of the plasma beta, the frequency, and the L-shell parameter, for different kappa values. In panel (a) (with $L=2$, $\bar{\omega}=0.1$), $\Lambda_c$ increases approximately linearly with the plasma beta in low-beta regimes, reaching $\Lambda_c \sim 0.42$ for cold plasmas. The increase is steeper for non-thermal plasmas. Panel (b) (with $L=2$, $\beta_0 = 0.1$ and $\kappa \geq 4$) shows a linear decrease of $\Lambda_c$ with the frequency, with a higher decline ratio for Maxwellian plasmas. For extremely non-thermal plasmas with $\kappa \sim 2$, $\Lambda_c$ decreases, reaching a minimum, then increases near $\omega/\Omega_{i0} \sim 0.3$.  Panel (c) (with $\bar{\omega} = 0.1$, $\beta_0 = 0.1$) shows a decrease of $\Lambda_c$ with the L-shell parameter, with stronger variation compared to plasma beta and frequency. Indeed $\Lambda_c$ can reach values of $\sim 0.1$ for $L\sim 10$, $\beta = 0.1$ and $\omega/\Omega_{i0} = 0.1$. However, since the dipole geomagnetic field model is generally more accurate near the plasmasphere, these results should be better adjusted for low $L$ values, depending on the planet. Also, the variation of $\Lambda_c$ with $L$ is almost independent of kappa; hence, only the Maxwellian case is displayed, as the curves for other $\kappa$ values overlap and are omitted for clarity.

Finally, to understand the underlying physics, we examine the forces involved in the equilibrium along geomagnetic field lines above and below $\Lambda_c$. Panels (a)-(b) of Figure \ref{fig:Plot_Forces} show the normalized forces that act in the equilibrium equation (Eq. \ref{eq:Force_balance_normalized}), as a function of the colatitude, for $\Lambda>\Lambda_c$ and $\Lambda < \Lambda_c$, respectively. Panel (c) shows a zoom over the gray-shaded region in panel (b). From inspecting Equation \eqref{eq:MMP_force_term}, since $\partial \varepsilon/\partial \mathcal{B} < 0$ for EMIC waves in low-beta regimes, the MMP force pushes the plasma against the gradient of the geomagnetic field, i.e, it tends to concentrate the plasma towards the equator. Consequently, in case of complex geometries, the MMP force might push the plasma to the local minima of the geomagnetic field, which is consistent with previous studies \citep{Lundin_Hultqvists_1989,Guglielmi_Lundin_2001,Nekrasov_2012,Nekrasov_Feygin_2016}. On the other hand, the wave amplitude, which is $\propto B/n^{1/4}$ for our WKB approximation, increases with colatitude due to the rising magnetic field magnitude and the decreasing density (pushed equatorward by the MMP force). Then, the spatial PF term, directed along the wave-amplitude gradient, drives plasma toward the planet's surface. Despite this, MMP dominates, as seen in panels (a) and (c) of Figure \ref{fig:Plot_Forces}, where the sum of the spatial and MMP terms of the normalized PF $\nu^2 \bar{f}$ is negative, directing plasma equatorward. For $\Lambda > \Lambda_c$, gravity is negligible near the equator compared to the PF as shown in panel (a) of Figure \ref{fig:Plot_Forces}. Therefore, the PF tendency to concentrate plasma towards the equator overcomes the gravity tendency to push it to the planet's surface, with the pressure equilibrating the PF. As colatitude increases, gravity increases in magnitude until it overcomes the PF, which decreases with colatitude (since it is inversely proportional to the magnetic field magnitude), thereby increasing density towards the surface. For $\Lambda < \Lambda_c$, gravity dominates everywhere, even closer to the equator, as shown in panel (c) of Figure \ref{fig:Plot_Forces}, pushing plasma planetward, balanced by the pressure. 

\section{Dependence on wave amplitude and average magnetospheric conditions}
\label{sec:7}

We analyze the magnitude of the ponderomotive effect dependence on the relative wave amplitude, $\nu$, and average conditions across the magnetospheres of Earth, Jupiter, and Saturn. Our analysis is grounded in statistical observational values of $\nu$ and plasma parameters, restricting the spatial domain to the dayside where EMIC waves are frequent and the magnetic field approximates a dipole. We focus on cold plasma regions ($\beta_0 < 0.3$), corresponding to the inner magnetospheres: $2.25 \leq L \leq 8$ (Earth), $5 \leq L \leq 10$ (Jupiter), and $3.6 \leq L \leq 7$ (Saturn). 

\begin{figure*}[ht]
\centering
\includegraphics[width=\linewidth]{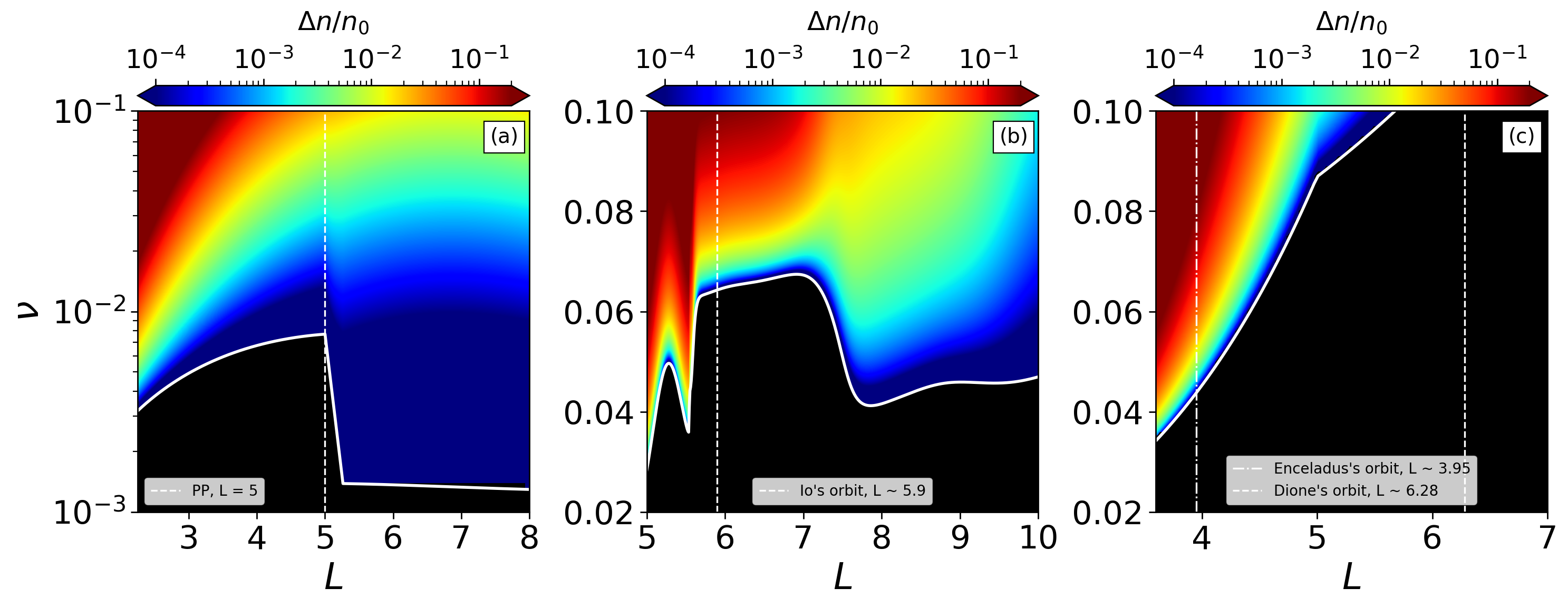}
\caption{Shaded isocontours of the relative difference between the equatorial and minimum densities, $\Delta n/n_0 = (n_0 - n_\text{min})/n_0$ (log scale), as a function of L-shell ($L$) and relative wave amplitude ($\nu$) under typical plasma conditions. (a) Earth's inner magnetosphere, where the white vertical dashed line denotes the plasmapause (PP). Here $\nu$ is denoted on a log scale. (b) Jupiter's inner magnetosphere, with Io's orbit marked by a white vertical dashed line. (c) Saturn's inner magnetosphere, with white vertical lines indicating the orbits of Enceladus (dash-dotted) and Dione (dashed). Across all panels, the white solid line represents the critical $\nu$ value associated with $\Lambda_c$, and the black-shaded regions indicate $\Delta n/n_0 = 0$, indicating that the density follows the standard profile with a single minimum at the equator.}
\label{fig:DensityMIN_colormap}
\end{figure*}

To evaluate the PF density redistribution, Figure \ref{fig:DensityMIN_colormap} displays the numerically computed $\Delta n/n_0 = (n_0 - n_\text{min})/n_0$ as a function of $L$ and $\nu$, where $n_\text{min}$ is the minimum density. This quantity indicates plasma accumulation at the equator due to the PF. Values $\Delta n/n_0 > 0.2$ are shown for context but exceed the validity of our perturbative model. We assume a Maxwellian velocity distribution for simplicity. We adopt $\bar{\omega}=0.1$ for Earth and $\bar{\omega}=0.3$ for Jupiter and Saturn, to account for the commonly observed EMIC frequency bands (see below). As detailed in Appendix \ref{sec:AppendixB}, which also outlines the empirical background profiles ($B_0$, $c_{A0}/c$, $\beta_0$) used for $n_{\text{min}}$ computation, the background gyrofrequency $\Omega_0$ is set to $1\,\Omega_{H^+}$ for Earth, $6.40 \times 10^{-2}\,\Omega_{H^+}$ for Jupiter, and $6.55 \times 10^{-2}\,\Omega_{H^+}$ for Saturn to account for ion composition.

\subsection{Earth's inner magnetosphere ponderomotive effect}

Figure \ref{fig:DensityMIN_colormap}.a shows density accumulation increases with wave amplitude. The L-shell dependence is governed by plasma beta, which counteracts the PF effect, and scales with the local Alfvén speed ratio as $\nu^2/C_g \propto (c_{A0}/c)^2$. Thus, $\Delta n/n_0$ decreases significantly at low $c_{A0}/c$ values. The abrupt variation near $L \sim 5$ reflects the plasmapause density gradient.

During quiet geomagnetic conditions, prevalent EMIC He$^+$ band emissions ($\omega < 0.25 \Omega_{H^+}$, mainly found at $4 \leq L \leq 6$ in prenoon and afternoon sectors) typically exhibit relative amplitudes of $\nu \sim 0.001 - 0.01$ \citep{Saikin_etal_2015,Usanova_etal_2014}. Consequently, the PF density perturbation is negligible ($\Delta n/n_0 < 10^{-4}$). However, extreme solar wind compressions can trigger high-amplitude Pc1 emissions ($\nu \sim 0.01 - 0.07$) across dawn and noon sectors spanning $L = 4 - 5.7$ \citep{Engebretson_etal_2015}, potentially leading to a non-negligible accumulation of plasma ($\Delta n/n_0 \sim 10^{-2}$).

We can extrapolate to higher L-shells ($L > 6$) to evaluate qualitative trends, though a rigorous treatment requires extending the model to account for non-dipolar geometries \citep{Antonova_Shabanskii_1968, Nekrasov_Feygin_2015}. In these outer regions, high-amplitude EMIC waves are commonly observed \citep{Min_etal_2012}. Dusk-sector emissions are dominated by the He$^+$ band ($L = 8-12$, $\nu \sim 0.01 - 0.1$), theoretically yielding a relative maximum of $\Delta n/n_0 \gtrsim 10^{-2}$. Conversely, dawn-sector waves are dominated by the H$^+$ band with lower amplitudes ($\nu \sim 0.01$) but higher normalized frequencies ($\omega \sim 0.5 \Omega_{H^+}$), approaching the proton cyclotron resonance. Since our current low-frequency formalism fails in this regime, this motivates future high-frequency model extensions, where we expect the PF to be stronger.

\subsection{Jupiter's inner magnetosphere ponderomotive effect}

Figure \ref{fig:DensityMIN_colormap}.b shows that Jupiter's inner magnetosphere requires high wave amplitudes ($\nu > 10^{-2}$) for the PF to redistribute plasma towards the equator. This results from low $c_{A0}/c$ values ($7 \times 10^{-4}$ to $2 \times 10^{-3}$) caused by massive heavy-ion mass-loading from the Io plasma torus \citep{Bagenal_1994,Saur_etal_2004}. Within this high-density region ($L \sim 5.9$), where EMIC waves are predominantly observed near the downstream region of the plasma torus, the strong background magnetic field restricts relative EMIC wave amplitudes to $\nu \lesssim 0.05$ \citep{Cao_etal_2025, Blanco_etal_2001}, effectively canceling the PF under typical conditions. However, substantial density redistribution ($\Delta n/n_0 \sim 10^{-1}$) can emerge during exceptionally high-amplitude events ($\nu > 0.08$) in the torus or further outwards ($L \sim 8$) before the local plasma beta increases. The observed wave power spectra are typically centered around the SO$^+$ gyrofrequency ($\Omega_{\text{SO}^+} \sim 0.33\,\Omega_0$) and extend up to the S$^+$ gyrofrequency ($\Omega_{\text{S}^+} \sim 0.49\,\Omega_0$) \citep{Blanco_etal_2001}; hence our choice of normalized frequency, $\bar{\omega} = 0.3$, appropriately represents the frequency bands of Jovian EMIC waves.

While our spatial domain is restricted to $L \leq 10$, where the low-beta plasma and dipole approximations hold \citep{Connerney_etal_1981}, the middle and outer magnetosphere ($L > 10$) present a physically interesting regime. Observations in these highly distorted, high-beta regions reveal dawnside EMIC waves with exceptionally large relative amplitudes of $\nu \sim 0.1 - 0.25$ \citep{Zhonghua_etal_2021,Yuan_etal_2024}. This suggests the Jovian outer magnetosphere frequently operates in a highly non-linear regime where the PF is expected to exert a substantial force against the background magnetic field gradient. A stronger PF interaction is also expected near heavy-ion resonances, strongly motivating future high-frequency model extensions.

\subsection{Saturn's inner magnetosphere ponderomotive effect}

Figure \ref{fig:DensityMIN_colormap}.c shows that Saturn's inner magnetosphere ($3.6 \leq L \leq 7$), where the magnetic field remains approximately dipolar \citep{Andre_etal_2008}, also requires high wave amplitudes ($\nu > 0.04$) for the PF to redistribute plasma. This environment features even lower $c_{A0}/c$ values ($1 \times 10^{-4}$ to $8 \times 10^{-4}$) than Jupiter due to weak magnetic fields and high concentrations of heavy water-group ions (O$^+$, OH$^+$, H$_2$O$^+$, H$_3$O$^+$) sourced by Enceladus's active cryovolcanism \citep{Meeks_etal_2016}. This, combined with the rapid increase of the plasma beta (see \ref{fig:Plot_parameters_Saturn}.c), demands higher $\nu$ values for the PF to act at larger L-shells, completely suppressing the PF redistribution for $L \gtrsim 6$ and $\nu < 0.1$. 

An equatorial statistical study by \citet{Meeks_etal_2016} shows continuous EMIC wave occurrence between the orbits of Enceladus and Dione, exhibiting a steep decay towards Rhea's orbit ($L = 8.74$) correlated with the radial decrease in water-group ion density. This study reports frequencies between $0.5 - 2.0\,\Omega_{\text{H}_2\text{O}^+}$ with a peak close to $\Omega_{\text{H}_2\text{O}^+}$, corresponding to roughly $0.42 - 1.70\,\Omega_0$. We therefore set our normalized frequency to $\bar{\omega} = 0.3$, which is at the upper limit of what our low-frequency approximation can hold. Observational amplitudes yield typical relative values of $\nu \sim 0.004 - 0.02$ \citep{Meeks_etal_2016,Long_etal_2022}. According to Figure \ref{fig:DensityMIN_colormap}.c, the PF effect is canceled under these typical conditions. However, significant density redistribution ($\Delta n/n_0 \gtrsim 10^{-1}$) could emerge during exceptional high-amplitude events ($\nu > 0.07$) near Enceladus's orbit ($L \sim 3.95$), where EMIC waves have their highest occurrence rate. In addition, since wave frequencies are frequently close to the water-group ion resonance, extending our framework to higher frequencies is strongly motivated.

\section{Conclusions}\label{sec:8}

We have analytically studied the effects of the non-thermal properties of velocity distributions observed in different planetary magnetospheres on the field-aligned plasma density redistribution induced by PF from traveling EMIC waves, using a dipole model of the geomagnetic field in a low-beta approximation. We have shown that the nonlinear plasma density solution due to PF in non-thermal plasmas resembles a second-kind phase transition, consistent with previous work \citep{Guglielmi_Hayashi_Lundin_Potapov_1999}. Furthermore, we calculated how the critical parameter $\Lambda_c$ for the phase transition varies with the plasma beta $\beta$, the kappa parameter $\kappa$, and the L-shell. For values above $\Lambda_c$, which would mainly depend on the wave amplitude responsible for the PF, plasma accumulation at the equator is predicted. This result is consistent with previous studies \citep{Allan_1993,Guglielmi_Hayashi_Lundin_Potapov_1999,Nekrasov_Feygin_2018}, which show accumulation at the geomagnetic field minimum, and specifically at the equator for a dipole geomagnetic model. 

As expected in low-beta plasmas, non-thermal and finite-temperature effects do not alter the qualitative behavior of plasma redistribution. Nevertheless, we found that the effects of the Kappa distribution and plasma beta counteract plasma accumulation at the geomagnetic field minimum due to the PF. We also demonstrated that this nonlinear effect may be relevant across planetary magnetospheres in the solar system, depending on the region and local plasma conditions. For instance, taking a low plasma beta, we found that for $\kappa = 2$, the equatorial density enhancement relative to the density minimum is less than $2\%$ compared to the $6\%$ in the Maxwellian case.

These results agree with \citet{allan_1992}, who, using a box-model simulation of the Earth's magnetosphere, showed that the nonlinear plasma density enhancement induced by PF from fast magnetosonic waves increases at lower local plasma beta. In the case of non-thermal plasmas considered here, the effect of the plasma beta is amplified by the Kappa distribution, reducing density enhancement. This result is especially relevant, given that measured kappa values in planetary magnetospheres of our solar system typically range from 2 to 10 in low-beta regions \citep{Schardt_1983, Krimigis_etal_1986, Mauk_1987, Krimigis_etal_1989, Mauk_etal_1996, Dialynas_etal_2009}, well within the validity of our approximation. We also found that the critical parameter $\Lambda_c$ for the phase transition increases linearly with plasma beta, with a steeper slope at low kappa. It decreases with frequency, though more slowly for small $\kappa$, and decreases with L-shell, with negligible dependence on $\kappa$.

In the present work, we employed the simplest geomagnetic model (dipole) to maintain generality and provide broad insights into non-thermal effects on plasma redistribution due to PF, results that can be qualitatively applied to different planetary magnetospheres of our solar system. However, each magnetosphere (and its different regions) has distinct structural characteristics that must be accounted for in specific studies. Despite this, because the MMP force associated with traveling EMIC waves pushes plasma against the magnetic field gradient, we expect this type of pulsation to contribute to plasma accumulation at magnetic minima in more complex geometries as well, in agreement with \citet{Nekrasov_Feygin_2018}. 

By applying this model to the dayside, low-beta inner magnetospheric regions of Earth, Jupiter, and Saturn, we demonstrated that the PF drives substantial equatorial density accumulation ($\Delta n/n_0 \sim 10^{-2} - 10^{-1}$) during exceptionally high-amplitude EMIC events ($\nu \sim 0.1$). These findings strongly motivate extending our framework to regions where the MHD approximation breaks down and non-thermal or multi-component plasma effects dominate. Specifically, the high-beta, collisionless environments of the terrestrial and Jovian outer magnetospheres frequently exhibit non-thermal distributions \citep{Kane_etal_1995, Pierrard_and_Stegen_2008, Kirpichev_etal_2021} and highly non-linear wave amplitudes ($\nu \gtrsim 10^{-1}$). Moreover, since standard time-averaged statistics underestimate the peak intensity of short wave-packets \citep{Shi_etal_2024}, this mechanism's impact likely exceeds current estimates. Additionally, future work will adapt this model for Mercury's magnetosphere, where ULF standing waves propagate near the ion gyrofrequency in Kappa-distributed plasmas \citep{Othmer_etal_1999, Christon_1987, Zhao_etal_2020, Harada_etal_2022}, and incorporate the multi-ion near-cyclotron resonances crucial for the heavy-ion-rich inner magnetospheres of Jupiter and Saturn.

\begin{acknowledgements}

We are grateful for the support of ANID Chile through the National Doctoral Scholarships No. 21231291 (JET), and FONDECYT grants Nos. 1230094 (FAA) and 1240281 (PSM).

\end{acknowledgements}

\bibliographystyle{aa}
\bibliography{References}

\appendix

\section{Ponderomotive force calculation}
\label{sec:AppendixA}

\subsection{Spatial term}

Using equations \eqref{eq:E_modulation} and \eqref{eq:varepsilon_approx} and neglecting terms at the second order of the plasma beta, we obtain that
\begin{equation}
\begin{split}
    \bar{f}_{(s)\parallel} &= \frac{1}{4}\left(\frac{c}{c_{A0}}\right)^{-1}\frac{(1-\bar{\omega})^{1/2}}{\varepsilon_c^{1/2}}\left\{-\frac{\alpha}{2}\left[\left(1+\frac{\delta_{\kappa 0}}{2}\right)-\frac{\delta_\kappa}{2}\right]\bar{\nabla}_\parallel \varepsilon_c \right. \\
    &\left. +\frac{\varepsilon_c \alpha}{2}\bar{\nabla}_\parallel \delta_\kappa + \varepsilon_c \left[\left(1+\frac{\delta_{0\kappa}}{2}\right)-\frac{\delta_\kappa}{2}\right]\bar{\nabla}_\parallel \alpha\right\}.
\label{eq:PFs_appendix1}
\end{split}
\end{equation}
Using equations \eqref{eq:varepsc_eq} and \eqref{eq:deltak_eq},  and considering that $\varepsilon_c \gg 1$ we obtain that
\begin{equation}
    \nabla_\parallel \varepsilon_c \approx \left(\frac{c}{c_{A0}}\right)^2\left\{T \bar{n}\nabla_\parallel\mathcal{B} + K\nabla_\parallel\bar{n}\right\},
\end{equation}

\begin{equation}
    \nabla_\parallel \delta_\kappa = \beta_{\kappa 0}\left\{S \bar{n}\nabla_\parallel \mathcal{B} + P \nabla_\parallel \bar{n}\right\},
\end{equation}

where $T = -(2\mathcal{B}-\bar{\omega})/[\mathcal{B}^2(\mathcal{B}-\bar{\omega})^2]$ and $S = (3/2)\bar{\omega}/(\mathcal{B}-\bar{\omega})^4$. Using the previous relations in Equation \eqref{eq:PFs_appendix1} we obtain that
\begin{equation}
\begin{split}
    \bar{f}_{(s)\parallel} &= \left(A_1+\beta_{\kappa 0} A_2\right)\bar{n}^{1/2} + \beta_{\kappa 0}A_3\bar{n}^{3/2} \\
    &+\left(A_4+\beta_{\kappa 0}A_5\right)\bar{n}^{-1/2}\bar{\nabla}_\parallel \bar{n} + \beta_{\kappa 0}A_6\bar{n}^{1/2}\bar{\nabla}_\parallel \bar{n},
\end{split}
\end{equation}
where
\begin{equation}
    A_1 = \frac{(1-\bar{\omega})^{1/2}}{4K^{1/2}}\left(-\frac{\alpha}{2}T\bar{\nabla}_\parallel \mathcal{B}+K\bar{\nabla}_\parallel \alpha\right), 
\end{equation}

\begin{equation}
    A_2 = \frac{(1-\bar{\omega})^{1/2}}{4K^{1/2}}\left(-\frac{\alpha}{2}GT\bar{\nabla}_\parallel \mathcal{B} + KG\bar{\nabla}_\parallel \alpha\right),
\end{equation}

\begin{equation}
    A_3 = \frac{(1-\bar{\omega})^{1/2}}{4 K^{1/2}}\left[\frac{\alpha}{2}\left(PT+KS\right)\bar{\nabla}_\parallel \mathcal{B} - \frac{KP}{2}\bar{\nabla}_\parallel \alpha\right], 
\end{equation}

\begin{equation}
    A_4 = -\alpha\frac{(1-\bar{\omega})^{1/2}}{8}K^{1/2},
\end{equation}

\begin{equation}
    A_5 = -\alpha\frac{(1-\bar{\omega})^{1/2}}{8}K^{1/2}G,
\end{equation}

\begin{equation}
    A_6 = \frac{3}{16}\alpha(1-\bar{\omega})^{1/2}PK^{1/2},
\end{equation}

where $G = -(1/4)\bar{\omega}/(1-\bar{\omega})^3$.

\subsection{MMP term}

Using \eqref{eq:E_modulation} and \eqref{eq:MMP_force_term} and neglecting terms at the second order of the plasma beta, we can deduce that the ponderomotive magnetic moment is given by

\begin{equation}
    M =  \frac{|E|^2}{16\pi B_0} \left(\frac{\partial \varepsilon_c}{\partial \mathcal{B}} - \varepsilon_c\frac{\partial \delta_\kappa}{\partial \mathcal{B}}\right).
\end{equation}
The partial derivatives are given by
\begin{equation}
    \frac{\partial \varepsilon_c}{\partial \mathcal{B}} = -\left(\frac{c}{c_{A0}}\right)^2 \frac{(2\mathcal{B}-\bar{\omega})}{\mathcal{B}^2(\mathcal{B}-\bar{\omega})^2}\bar{n} = \left(\frac{c}{c_{A0}}\right)^2T\bar{n},
\end{equation}

\begin{equation}
    \frac{\partial \delta_\kappa}{\partial \mathcal{B}} = \frac{3}{2}\beta_{\kappa 0}\frac{\bar{\omega}}{(\mathcal{B}-\bar{\omega})^4}\bar{n} = \beta_{\kappa 0}S\bar{n}.
\end{equation}
Therefore, we have that the magnetic moment pumping along the magnetic field lines $f_{(MMP)\parallel} = B_0 R_p M \bar{\nabla}_\parallel \mathcal{B}$ is given by
\begin{equation}
    f_{(MMP)\parallel} = \left(A_7 + \beta_{\kappa 0}A_8\right)\bar{n}^{1/2} + \beta_{\kappa 0}A_9\bar{n}^{3/2},
\end{equation}
where,
\begin{equation}
    A_7 = \frac{(1-\bar{\omega})^{1/2}}{4K^{1/2}}\alpha T\bar{\nabla}_\parallel \mathcal{B},
\end{equation}

\begin{equation}
    A_8 = \frac{(1-\bar{\omega})^{1/2}}{4K^{1/2}}\alpha TG\bar{\nabla}_\parallel \mathcal{B},
\end{equation}

\begin{equation}
    A_9 = \frac{(1-\bar{\omega})^{1/2}}{4K^{1/2}}\left(\frac{PT}{2}-KS\right)\alpha\bar{\nabla}_\parallel \mathcal{B}.
\end{equation}

\subsection{Terms in the ODE}

We can calculate explicitly the terms $\Phi_i$ (we do it for $\alpha = 1$) involved in the ODE given by Equation \eqref{eq:ODE},

\begin{equation}
\begin{split}
    \Phi_1(x_\parallel,\beta_{\kappa 0}) =& -\frac{(1-\bar{\omega})^{1/2}}{8}\left(1-\beta_{\kappa 0}\frac{\bar{\omega}}{4(1-\bar{\omega})^3}\right) \\
    &\times \frac{(2\mathcal{B}(x_\parallel)-\bar{\omega})}{\mathcal{B}(x_\parallel)^{3/2}(\mathcal{B}(x_\parallel)-\bar{\omega})^{3/2}}\bar{\nabla}_\parallel\mathcal{B}(x_\parallel),
\end{split}
\end{equation}

\begin{equation}
    \Phi_2(x_\parallel,\beta_{\kappa 0}) = \beta_{\kappa 0}\frac{\bar{\omega}(1-\bar{\omega})^{1/2}}{8}\frac{\left(\frac{1}{2}\mathcal{B}(x_\parallel)-\bar{\omega}\right)}{\mathcal{B}(x_\parallel)^{3/2}(\mathcal{B}(x_\parallel)-\bar{\omega})^{9/2}}\bar{\nabla}_\parallel \mathcal{B}(x_\parallel),
\end{equation}

\begin{equation}
   \Phi_3(x_\parallel,\beta_{\kappa 0}) = \frac{1}{8}\sqrt{\frac{1-\bar{\omega}}{\mathcal{B}(x_\parallel)(\mathcal{B}(x_\parallel)-\bar{\omega})}}\left(1-\beta_{\kappa 0}\frac{\bar{\omega}}{(1-\bar{\omega})^3}\right),
\end{equation}

\begin{equation}
    \Phi_4(x_\parallel,\beta_{\kappa 0}) = \beta_{\kappa 0}\frac{3}{32}\frac{(1-\bar{\omega})^{1/2}\bar{\omega}}{\mathcal{B}(x_\parallel)^{1/2}(\mathcal{B}(x_\parallel)-\bar{\omega})^{7/2}}.
\end{equation}

\section{Empirical estimation of magnetospheric plasma parameters}
\label{sec:AppendixB}

\subsection{Earth}

\begin{figure*}[ht]
\centering
\includegraphics[width=\linewidth]{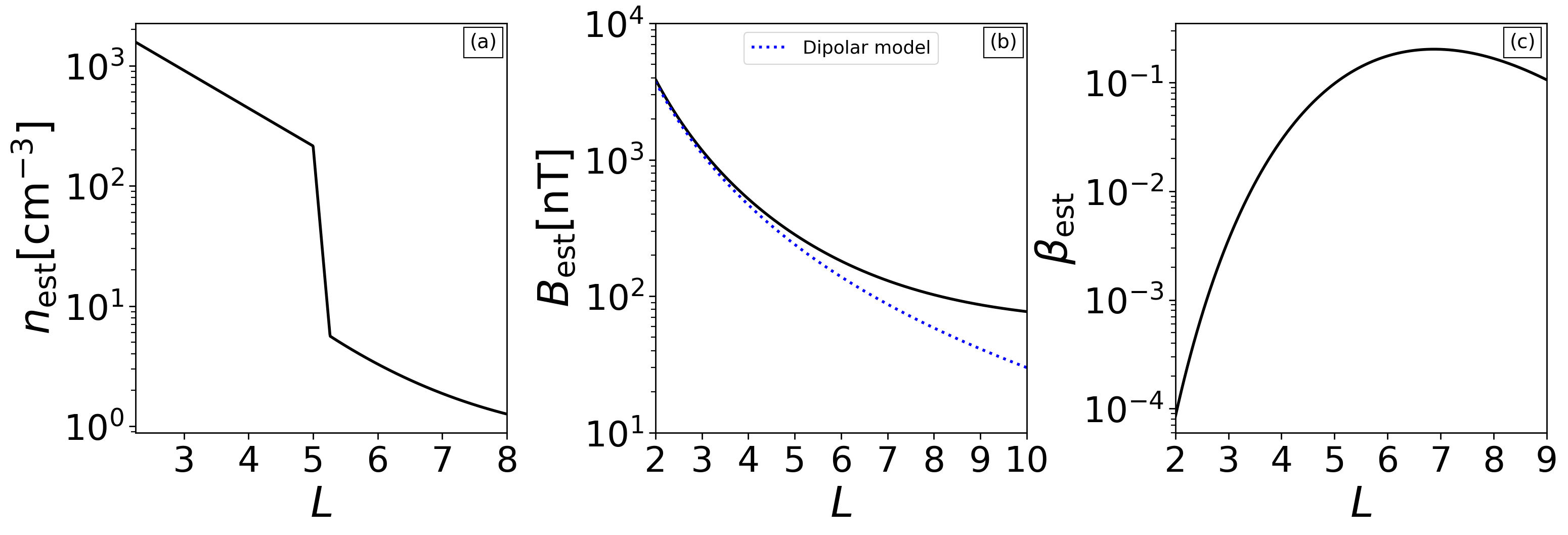}
\caption{Empirical estimates of the average equatorial plasma parameters in Earth's dayside magnetosphere as a function of L-shell. Panel (a): Estimated plasma density, $n_{\text{est}}$ ($\text{cm}^{-3}$).Panel (b): Estimated magnetic field magnitude, $B_{\text{est}}$ (black solid line, $\text{nT}$), compared to the standard terrestrial dipole model (blue dotted line).Panel (c): Estimated plasma beta, $\beta_\text{est}$.}
\label{fig:Plot_Parameters_Earth}
\end{figure*}

To obtain an estimated average equatorial plasma density profile $n_{\text{est}}$ in Earth's magnetosphere, we use \citet{Carpenter_and_Anderson_1992} empirical model, derived from ISEE 1 satellite measurements and whistler data. The model is valid for the range $2.25<L<8$ and $00\text{-}15 \text{ MLT}$ under relatively steady global magnetic conditions. It is defined piecewise for the "saturated" plasmasphere ($2.25 \leq L \leq L_{\text{ppi}}$), the plasmapause ($L_\text{ppi} <L < L_{\text{ppo}}$), and the plasma trough ($L_\text{ppo} < L \leq 8$)

\begin{equation}
    \frac{n_{\text{est}}(L)}{\text{cm}^{-3}} = \left\{\begin{array}{ll}
    10^{-(0.3145L-3.9043)} &, 2.25 \leq L \leq L_{\text{ppi}} \\
    & \\
    n_0(L_{\text{ppi}}) 10^{-\frac{(L-L_{\text{ppi}})}{(0.1+0.011(t-6))}} &, L_{\text{ppi}} \leq L \leq L_{\text{ppo}} \\
    & \\
    (5800 + 300t)\left(\frac{L}{L_{\text{ppo}}}\right)^{-4.5} &\\
    + \left(1-e^{-(L-2)/10}\right) &, L_{\text{ppo}} \leq L \leq 8
  \end{array} \right.
\label{eq:EstimatedDensity_Earth}
\end{equation}
For this analysis, we set $t=12$ (noon) and $L_\text{ppi} = 5$, consistent with typical plasmaspheric observations where $L_{\text{ppi}}$ extends up to $\sim 4\text{-}5 R_E$. The outer boundary $L_\text{ppo}$ of the plasmapause, a sharp density gradient region typically narrower than $0.5 L$ \citep{Kwon_etal_2015}, is determined by demanding continuity
\begin{equation*}
n_0(L_{\text{ppi}})10^{-\frac{(L_{\text{ppo}}-L_{\text{ppi}})}{(0.1+0.011(t-6))}} = (5800 + 300t) + \left(1-e^{-(L_{\text{ppo}}-2)/10}\right).
\end{equation*}
This profile is shown in Figure \ref{fig:Plot_Parameters_Earth}.a.

For the estimated magnetic field magnitude $B_{\text{est}}$, we consider \citet{Mead_1964} distorted dipole model, which accounts for the dayside compression of the geomagnetic field by solar wind pressure. The equatorial magnetic field magnitude (in nT) at noon is
\begin{equation}
\frac{B_{\text{est}}(L)}{\text{nT}} = \left(\frac{3.1}{L^3} + \frac{2.515}{r_b^3} + \frac{\sqrt{3}\times 1.215}{r_b^4}L\right) \times 10^{4},
\label{eq:EstimatedMagneticField_Earth}
\end{equation}
where $r_b = 10$ is the adopted typical normalized distance (in $R_E$) to the magnetopause \citep{Cahill_and_Patel_1967}. Because this model incorporates surface currents at the magnetopause boundary, the field decays more slowly than a pure dipolar field $B_{\text{dipolar}} \sim (3.1/L^3)\times 10^4\text{nT}$, as compared in Figure \ref{fig:Plot_Parameters_Earth}.b.

Using Equations \ref{eq:EstimatedDensity_Earth} and \ref{eq:EstimatedMagneticField_Earth}, the estimated Alfvén velocity $c_{\text{A,est}}$ applied in section \ref{sec:7} for the Earth's magnetosphere is
\begin{equation}
\frac{c_{\text{A,est}}}{c} = \frac{B_{\text{est}}}{\sqrt{4\pi M n_{\text{est}}}} \sim \frac{(B_\text{est}/\text{nT})}{\sqrt{(M/m_\text{p})(n_{\text{est}}/\text{cm}^{-3}})} \times \left(7.27 \times 10^{-5} \right)
\label{eq:AlfvenVelocityAppendix}
\end{equation}
where $m_p$ is the proton mass and $M$ the ion mass. Since protons mainly populate Earth's magnetosphere, we can consider $M = m_{\text{p}}$. Based on these estimations $c_{\text{A,est}}/c \sim 10^{-3}\text{ - }8\times10^{-3}$ for all $L$-shell values considered ($2.25 \leq L \leq 8$), consistent with standard observations \citep{Sarris_etal_2009}.

To estimate average thermal pressure, we use AMPTE/CCE-CHEM proton distribution data averaged over 1985-1987 \citep{Michelis_etal_1999}. We extracted perpendicular and parallel proton pressures (in nPa) via WebPlotDigitizer \citep{Rohatgi_2024} and fitted them to a gamma-type empirical function, $P(L)/\text{nPa} = AL^a\exp^{-bL}$. The resulting best-fit parameters $(A,b,a)$ are $(0.100,7.466,1.688)$ for the perpendicular component $P_{\perp,\text{est}}$, and $(0.047,7.431,1.617)$ for the parallel component $P_{\parallel,\text{est}}$. The total isotropic pressure $P_{\text{est}} = \frac{1}{3}\left(P_{\parallel,\text{est}} + 2P_{\perp,\text{est}}\right)$ yields an average plasma beta of
\begin{equation}
\beta_{\text{est}} = \frac{8\pi(P_{\text{est}}/\text{nPa})}{(B_{\text{est}}/\text{nT})^2} \times 10^2.
\end{equation}
As shown in Figure \ref{fig:Plot_Parameters_Earth}.c, $\beta_\text{est} \sim 0.1$ near the magnetopause and drops below $10^{-3}$ inside the plasmasphere.

\subsection{Jupiter}

To obtain an estimated average equatorial plasma density profile in the inner Jovian magnetosphere, we consider the observations reported by \citet{Bagenal_1994}. This study presents a description of the plasma conditions in the Io plasma torus, between $L = 5$ and $L = 10 R_J$, based on Voyager 1 data collected in March 1979. In their analysis, the plasma characteristics observed along the spacecraft trajectory were extrapolated along magnetic field lines by numerically solving the diffusive equilibrium equations. This method yielded radial profiles of plasma properties at the centrifugal equator, as well as meridian-plane density maps for the major ionic species. 

Using WebPlotDigitizer \citep{Rohatgi_2024}, we extracted the radial density profiles (in $\text{cm}^{-3}$) for each ionic species at the centrifugal equator as presented in that study. We then fit these profiles on a logarithmic scale using a combination of a Gaussian function (representing the torus source of newborn ions) and two additional functions that capture the slope of the density decrease as $L$ increases. Thus, for each species $\alpha$, the estimated density is given by $n_{\text{est},\alpha}(L) = f_\text{core}(L) + [1-\sigma_1(L)]\sigma_2(L)f_{\text{ledge}}(L) + [1-\sigma_2(L)]f_\text{tail}(L)$, where
\begin{equation*}
\begin{split}
    &f_\text{core}(L) = A_c e^{-(L-L_c)^2/2D_c^2}, \\
    &f_\text{ledge}(L) = A_1 10^{-b_1 L }, \\
    &f_\text{tail}(L) = A_2 10^{-b_2 L }, \\
    &\sigma_1(L) = \left(1+e^{2(L-L_1)/D_1}\right)^{-1}, \\
    &\sigma_2(L) = \left(1+e^{2(L-L_2)/D_2}\right)^{-1}. \\
\end{split}
\end{equation*}

\begin{figure*}[ht]
\centering
\includegraphics[width=\linewidth]{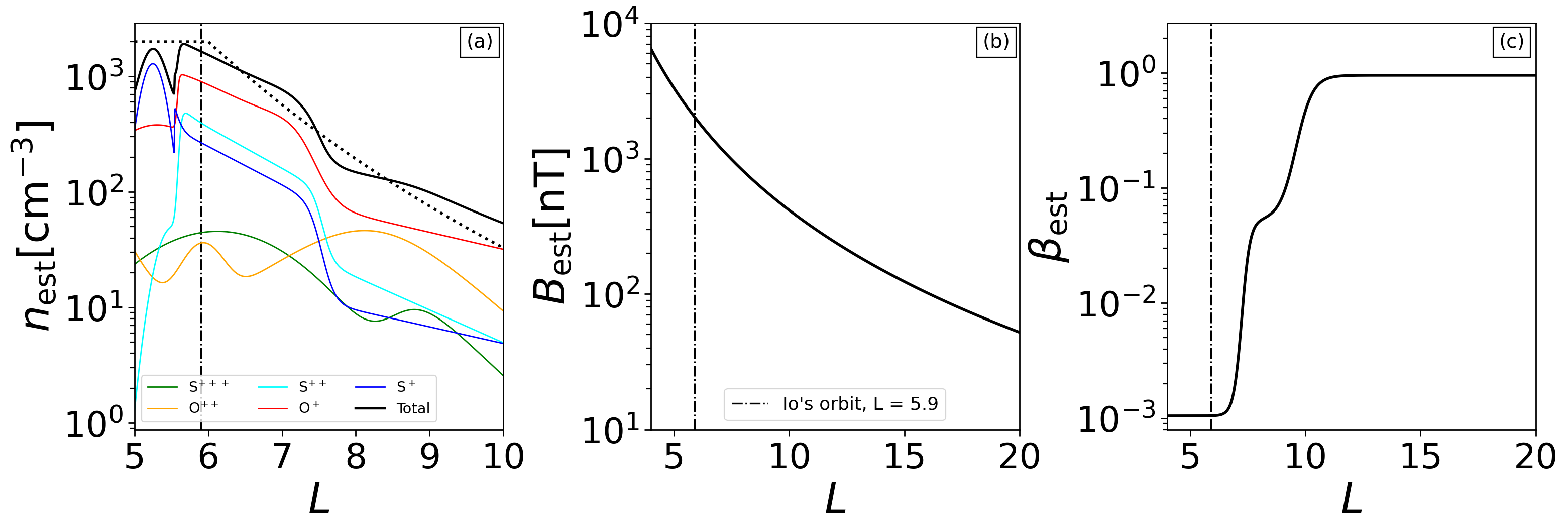}
\caption{Empirical estimates of the average equatorial plasma parameters in Jupiter's dayside magnetosphere as a function of L-shell. Panel (a) Estimated plasma density, $n_{\text{est},\alpha}$ ($\text{cm}^{-3}$) for each species $\alpha$. Solid lines denote the major ion species: O$^+$ (red), O$^{++}$ (orange), S$^+$ (blue), S$^{++}$ (cyan), S$^{+++}$ (green), and the total density (black). The empirical density model by \citet{Bagenal_etal_2011} is shown as a black dotted line. Panel (b): Estimated magnetic field magnitude, $B_{\text{est}}$ ($\text{nT}$). Panel (c): Estimated plasma beta, $\beta_\text{est}$. Across all panels, the black dash-dotted line indicates Io's orbit at $L = 5.9$.}
\label{fig:Plot_Parameters_Jupiter}
\end{figure*}

The best-fit values for each species are listed in Table REFTABLE, and the corresponding estimated average density profiles are shown in Figure \ref{fig:Plot_Parameters_Jupiter}.a. In this cold region of the inner torus, the composition is dominated by S$^+$ and O$^+$ ions, with significant contributions from S$^{++}$, S$^{+++}$, and O$^{++}$. Furthermore, an upper limit of 0.5\% can be placed on the proton density in the cold torus \citep{Bagenal_1994}, based on the portion of the distribution extending above 10~eV. This makes protons a minor constituent of the ion plasma in this inner region. However, in the middle magnetosphere, protons comprise a significant fraction (20--50\%) of the total plasma \citep{McNutt_etal_1981}. 

We compare the total density predicted by our estimation (Figure \ref{fig:Plot_Parameters_Jupiter}.a, black solid line) with the empirical model by \citet{Bagenal_etal_2011}, which combines Voyager and Galileo plasma sheet measurements from the outer boundary of Io's orbit ($L \sim 6$) to $L \sim 50$
\begin{equation}
    \frac{n_{\text{est}}(L)}{\text{cm}^{-3}} = 1987 \left(\frac{L}{6}\right)^{-8.2} + 14 \left(\frac{L}{6}\right)^{-3.2} + 0.05 \left(\frac{L}{6}\right)^{-0.65}.
\end{equation}
This model is depicted by the dotted black line in Figure \ref{fig:Plot_Parameters_Jupiter}.a. Both profiles exhibit strong consistency.

For the magnetic field magnitude, we assume a standard dipole model
\begin{equation}
    B_{\text{est}}(L) = \frac{4.17\times 10^5}{L^3} \,\text{nT}.
\end{equation}
This approximation remains valid in the inner Jovian magnetosphere up to $L \sim 10$, where contributions from currents flowing in the magnetospheric equatorial plasma sheet (the magnetodisc) are negligible \citep{Connerney_etal_1981, Khurana_etal_2004}. The corresponding profile is shown in Figure \ref{fig:Plot_Parameters_Jupiter}.b.

Because our framework assumes a single-ion species, we must self-consistently adapt our equations to this multi-species environment to account for the different mass populations. We achieve this by defining an effective ion mass $M_\text{eff}$ and an effective charge $q_\text{eff}$ as follows
\begin{equation}
    M_{\text{eff}} = \frac{1}{n_{\text{est,Total}}}\sum_\alpha n_{\text{est},\alpha} m_\alpha,
\end{equation}
\begin{equation}
    q_{\text{eff}} = \frac{1}{n_{\text{est,Total}}}\sum_\alpha n_{\text{est},\alpha} q_\alpha,
\end{equation}
where $n_{\text{est,Total}} = \sum_\alpha n_{\text{est},\alpha}$. Averaging these quantities over the considered L-shell range (denoted by $\langle \rangle$) yields an average effective mass $\langle M_\text{eff} \rangle$ and charge $\langle q_{\text{eff}}\rangle$. By substituting $M = \langle M_\text{eff} \rangle \sim 21.831\,m_p$ into Equation \ref{eq:AlfvenVelocityAppendix}, we estimate the Alfvén velocity for the Jovian magnetosphere. Based on these calculations, we obtain $c_{A,\text{est}}/c \sim 7 \times 10^{-4} - 2 \times 10^{-3}$ across the evaluated range ($5 \leq L \leq 10$).

Consequently, the ion gyrofrequency $\Omega_0$ used to normalize the wave frequency is no longer the proton gyrofrequency $\Omega_{H^+}$ (as is the case on Earth). Instead, it must be adjusted to $\Omega_0 \sim \left[\left(\langle q_\text{eff}\rangle/e\right) / \left(\langle M_{\text{eff}}\rangle/m_p\right)\right] \Omega_{H^+} \sim 6.40 \times 10^{-2} \Omega_{H^+}$, given that $\langle q_\text{eff}\rangle \sim 1.397\, e$, with $e$ representing the elementary charge.

Finally, to estimate the average plasma beta, we rely on bulk plasma parameters derived from Voyager 1 and 2 data \citep{Mauk_etal_1996}, covering the vicinity of the Io plasma torus, the inner magnetosphere, and a portion of the middle magnetosphere ($5 \leq L \leq 20$). We extracted these plasma beta values using WebPlotDigitizer \citep{Rohatgi_2024} and fitted them to a double-sigmoid empirical function
\begin{equation}
\log_{10}(\beta_\text{est}(L)) = C + \frac{A_1}{1+e^{-k_1(L-x_1)}} + \frac{A_2}{1+e^{-k_2(L-x_2)}}.
\end{equation}
The resulting best-fit parameters $(C, A_1, k_1, x_1, A_2, k_2, x_2)$ are $(-2.979, 1.696, 5.633, 7.242, 1.262, 2.748, 9.593)$.

\subsection{Saturn}

To estimate the average equatorial plasma density profile in Saturn's inner magnetosphere, we rely on electron and ion density measurements from the Cassini RPWS and CAPS instruments, collected between June 2004 and September 2007 \citet{Persoon_etal_2009}. The observations cover latitudes from the equator to $35^{\circ}$ and radial distances $3.6 \leq L \leq 10$. This inner region is heavily populated by water-group ions ($W^+$) originating from the plumes of Enceladus' southern polar region, alongside hydrogen ions ($H^+$).

At lower $L$ values, ion densities increase radially toward a broad equatorial peak, remaining approximately constant in the $4 < L \leq 5$ range. The equatorial density profile for $W^+$ reaches a maximum of $62\,\text{cm}^{-3}$ at $L = 4.8$, while $H^+$ peaks at $9\,\text{cm}^{-3}$ at $L = 4.6$. Beyond $L = 5$, both species diffuse radially outward, exhibiting inverse $L$-shell dependencies: $n_{\text{eq},W^+} \propto L^{-4.3}$ and $n_{\text{eq},H^+} \propto L^{-3.2}$.

Using these observations, we approximate the total estimated number density $n_{\text{est,Total}}$ for $4 < L \leq 10$ by assuming a constant peak density up to $L=5$ and subsequently applying the respective decay rates:
\begin{equation}
    \frac{n_{\text{est,Total}}(L)}{\text{cm}^{-3}} = 
    \begin{cases}
        59 & 4 < L \leq 5 \\
        n_{\text{est},W^+}(L) + n_{\text{est},H^+}(L) & 5 < L \leq 10
    \end{cases}
\end{equation}
where $n_{\text{est},W^+}(L) = 52 (L/5)^{-4.3}$ and $n_{\text{est},H^+}(L) = 7 \left(L/5\right)^{-3.2}$ are the estimated density profiles for water-group ions and protons, respectively. While more comprehensive models are necessary for extended radial distances \citep{Bagenal_etal_2011}, this piecewise function sufficiently captures the density profile required for our analysis in the inner magnetosphere. 

\begin{figure*}[ht]
\centering
\includegraphics[width=\linewidth]{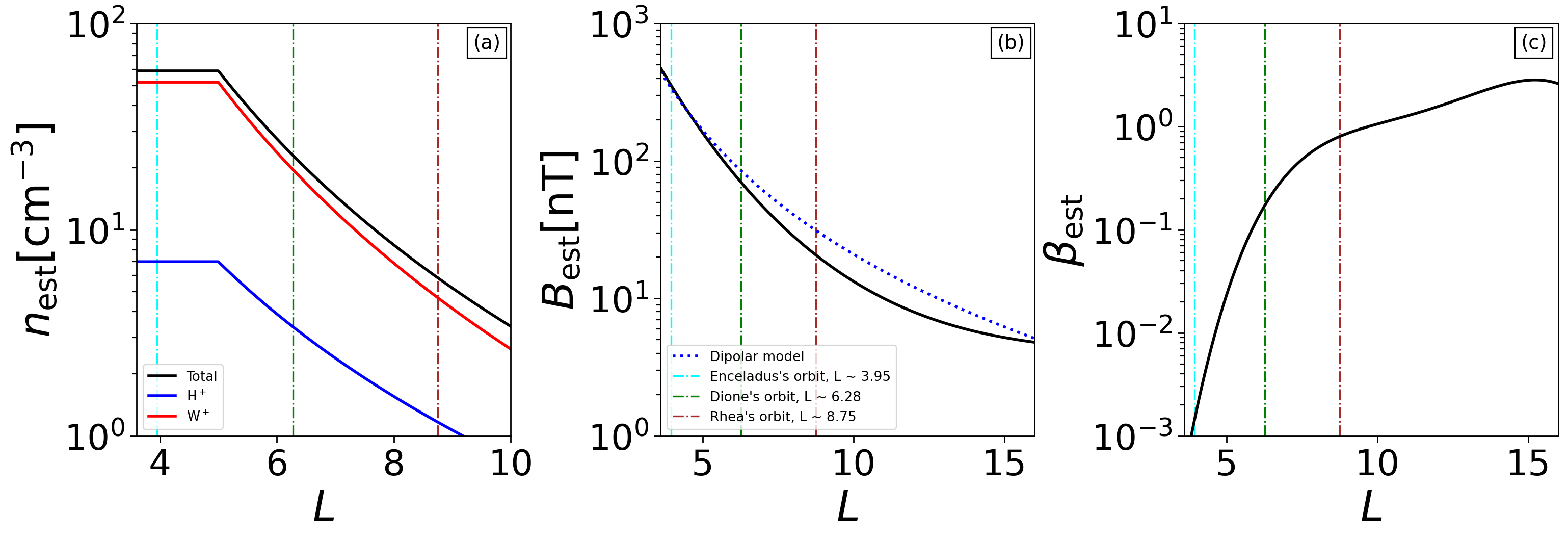}
\caption{Empirical estimates of the average equatorial plasma parameters in Saturn's dayside magnetosphere as a function of L-shell. Panel (a): Estimated plasma density, $n_{\text{est},\alpha}$ ($\text{cm}^{-3}$) for each species $\alpha$. Solid lines denote the specific ion species: W$^+$ (red), H$^+$ (blue), and total density (black). Panel (b) Estimated magnetic field magnitude, $B_{\text{est}}$ (black solid line, $\text{nT}$), compared to the standard Saturnian dipole model (blue dotted line). Panel (c) Estimated plasma beta, $\beta_\text{est}$. Across all panels, dash-dotted lines mark the orbits of the main moons: Enceladus (cyan, $L \sim 3.95$), Dione (green, $L \sim 6.28$), and Rhea (brown, $L \sim 8.75$).}
\label{fig:Plot_parameters_Saturn}
\end{figure*}

As shown in Figure \ref{fig:Plot_parameters_Saturn}.a, Saturn's inner magnetosphere is strongly dominated by heavy water-group ions ($W^+$). To self-consistently adapt our single-ion equations to this multi-species environment, we define an average effective ion mass $M_{\text{eff}} = \langle (m_p n_{\text{est},H^+} + 18 m_p n_{\text{est},W^+})/n_{\text{est,Total}} \rangle \sim 15.262 \,m_p$ and an effective charge $q_\text{eff} = 1$. Consequently, the adjusted equatorial ion gyrofrequency becomes $\Omega_{i0} \sim 6.55 \times 10^{-2}\, \Omega_{H^+}$.

Finally, to estimate the magnetic field magnitude and average plasma beta, we rely on the study by \citet{Sergis_etal_2017}, which analyzes particle and magnetic field data collected between July 2004 and December 2013 by the Cassini spacecraft in Saturn's equatorial magnetosphere, covering radial distances from $5$ to $16 R_S$.

They provide empirical polynomial fits for the dayside pressure (in Pa) and the magnetic field $z$-component (in nT). Although these fits are officially valid for $5 < L < 16$, we extend their application inward up to $L = 4$:
\begin{equation}
    \log_{10}P_{\text{est}}(L) = a_0 + a_1 L + a_2 L^2 + a_3 L^3 + a_4 L^4,
\end{equation}
\begin{equation}
    \log_{10}B_{z,\text{est}}(L) = d_0 + d_1 L + d_2 L^2 + d_3 L^3,
\end{equation}
where $\left(a_0,a_1,a_2,a_3,a_4\right) = \left(-17.96,3.73,-0.5697,0.0353\right.$ $ \left.-0.0007795\right)$ and $\left(d_0,d_1,d_2, d_3\right) = \left(4.338,-0.5568,0.0286,-0.0005052\right)$. The fitted component $B_{z}$ corresponds to the off-equator (north-south) magnetic field. However, because their analysis was restricted to measurements where $|B_r| < 0.5 \,\text{nT}$ and $|B_r| < 0.2 |B|$, we can safely assume $B_\text{est} \sim B_{z,\text{est}}$. As illustrated in Figure \ref{fig:Plot_parameters_Saturn}.b, this field decays faster than a pure dipolar field, $B_{\text{dipolar}} \sim (2.1 \times 10^4/L^3)\,\text{nT}$. This accelerated decay is driven by the centrifugal expansion of the trapped plasma, which forms a magnetodisc and drives an equatorial ring current whose induced magnetic field opposes the planetary dipole in the inner and middle magnetosphere \citep{Connerney_etal_1983,Arridge_etal_2008}. Based on these estimations, we obtain $c_{A,\text{est}}/c \sim 1 \times 10^{-4} - 8 \times 10^{-4}$ across the evaluated range ($3.6 \leq L \leq 10$).

Figure \ref{fig:Plot_parameters_Saturn}.c presents the estimated equatorial plasma beta profile as a function of $L$. The plasma beta increases with radial distance, remaining $> 1$ beyond $L \sim 8\text{--}10 R_S$ across all local times \citep{Sergis_etal_2017}, and reaches peak values of $\beta \sim 3$, $4$, and $6\text{--}8$ for the dayside, dawn, and dusk/nightside sectors, respectively. Conversely, moving toward the inner magnetosphere, the magnetic field strength increases sharply with decreasing radial distance. This causes the plasma beta to drop rapidly, becoming $< 0.1$ at $L = 5 R_S$ for all local times. On the dayside, $\beta < 0.3$ for $L < 7$. Since our theoretical framework requires a low-beta regime, this restricted interval is the primary focus of our analysis.

\clearpage

\renewcommand{\tabcolsep}{5pt}
\begin{table*}[hp]
\centering
\caption{Best-fit values for the estimated density profile in the Jovian magnetosphere for each species. }
\label{tab:JupiterMagnetosphere_BestFitValues}
\begin{tabular}{cccccccccccc}
\toprule
Species & $A_c$ & $L_c$ & $D_c$ & $L_1$ & $D_1$ & $L_2$ & $D_2$ & $A_1$ & $b_1$ & $A_2$ & $b_2$  \\ 
\midrule
\addlinespace[3pt]
O$^+$ & 3.810 (2) & 5.306 & 6.453 (-1) & 5.586 & 2.626 (-2) & 7.325 & 2.947 (-1) & 4.330 (3) & 1.389 (-1) & 9.437 (2) & 1.471 (-1)  \\
\addlinespace[3pt]
S$^+$ & 1.290 (3) & 5.248 & 1.520 (-1) & 5.541 & 2.949 (-3) & 7.462 & 1.718 (-1) & 2.429 (4) & 3.319 (-1) & 1.341 (2) & 1.440 (-1)  \\
\addlinespace[3pt]
S$^{++}$ & 4.817 (1) & 5.497 & 1.841 (-1) & 5.614 & 4.266 (-2) & 7.500 & 1.440 (-1) & 4.799 (4) & 3.538 (-1) & 3.467 (3) & 2.846 (-1)  \\
\addlinespace[3pt]
O$^{++}$ & 2.523 (1) & 5.922 & 2.492 (-1) & 1.139 (-3) & 2.253 (-3) & 8.247 & 8.831 (-1) & 1.384 (8) & 1.342 & 4.001 (6) & 5.626 (-1)  \\
\addlinespace[3pt]
S$^{+++}$ & 4.561 (1) & 6.125 & 9.810 (-1) & - & - & 8.676 & 4.185 (-1) & 0 & - & 2.200 (6) & 5.936 (-1)  \\
\bottomrule
\end{tabular}
\tablefoot{The numbers in parentheses represent the magnitude order.}
\end{table*}

\end{document}